\documentclass[preprint2]{aastex}
\usepackage{graphics,graphicx}
\usepackage{color}
\usepackage{epsf}
\usepackage{epsfig}
\usepackage{dcolumn}

\newcommand{\xmm}{{\em XMM-Newton}}
\newcommand{\spitz}{{\em Spitzer}}
\newcommand{\hii}{H\,{\sc ii}}
\newcommand{\wo}{WR\,142}
\newcommand{\ber}{Berkeley\,87}

\newcommand{\Lx}{\mbox{$L_{\rm X}$}}
\newcommand{\Lsun}{\mbox{$L_\odot$}}
\newcommand{\Msun}{\mbox{$M_\odot$}}

\newcommand{\myr}{\mbox{$M_\odot\,{\rm yr}^{-1}$}}
\newcommand{\lsim}{\raisebox{-.4ex}{$\stackrel{<}{\scriptstyle \sim}$}}
\newcommand{\msim}{\raisebox{-.4ex}{$\stackrel{>}{\scriptstyle \sim}$}}
\newcommand{\mim}{\mbox{$\mu$m}}

\def \etal   {\hbox{et~al.\/}}
\def\changed{}
\def\nchanged{}
\def \new{}
\usepackage{txfonts}

\shorttitle{Hard diffuse X-ray emission in the star-forming 
region ON\,2: discovery with \xmm.}
\shortauthors{Oskinova et al.}
\begin{document}
\title{\new{Hard X-ray emission in the star-forming 
region ON\,2: discovery with \xmm.}}

\author{L.~M.~Oskinova}
\affil{Institute for Physics and Astronomy, University of Potsdam,
14476 Potsdam, Germany\\
\email{lida@astro.physik.uni-potsdam.de}}
\author{R.~A.~Gruendl}
\affil{Department of Astronomy, University of Illinois,
1002 West Green Street, Urbana, IL 61801, USA}
\author{R.~Ignace}
\affil{Department of Physics and Astronomy, East Tennessee State University,
Johnson City, TN 37614, USA}
\author{Y.-H.~Chu}
\affil{Department of Astronomy, University of Illinois,
1002 West Green Street, Urbana, IL 61801, USA}
\author{W.-R. Hamann}
\affil{Institute for Physics and Astronomy, University of Potsdam,
14476 Potsdam, Germany}
\author{A.~Feldmeier}
\affil{Institute for Physics and Astronomy, University of Potsdam,
14476 Potsdam, Germany}

\begin{abstract} 

We obtained X-ray \xmm\ observations of the open cluster \ber\ and the
massive star-forming region (SFR) ON\,2. In addition, archival infrared
{\it Spitzer Space Telescope} observations were used to study the
morphology of ON\,2, to uncover young stellar objects, and to
investigate their relationship with the X-ray sources. It is likely 
that the SFR ON\,2 and \ber\ are at the same distance, 1.23 kpc, and 
hence are associated.  The \xmm\ observations detected X-rays from
massive stars in \ber\ as well as diffuse emission from the SFR ON\,2.
The two patches of diffuse X-ray emission are encompassed in the
shell-like \hii\ region GAL\,75.84+0.40 in the northern part of ON\,2
and in the ON\,2S region in the southern part of ON\,2.  The diffuse
emission from GAL\,75.84+0.40 suffers an absorption column equivalent to
$A_{\rm V}\approx 28$\,mag.  Its spectrum can be fitted either with a
thermal plasma model at $T\msim 30$\,MK\, or by an absorbed power-law
model with $\gamma$$\approx$$-2.6$. The X-ray luminosity of
GAL\,75.84+0.40 is $L_{\rm X}$$\approx$$1\times 10^{32}$\,erg s$^{-1}$.
The diffuse emission from ON\,2S is adjacent to the ultra-compact \hii\
(UC\hii) region Cygnus\,2N, but does not coincide with it or with any
other known UC\hii\ region.  It has a luminosity of $L_{\rm
X}$$\approx$$6\times 10^{31}$\,erg s$^{-1}$.  The spectrum can be fitted
with an absorbed power-law model with $\gamma$$\approx$$-1.4$.  We adopt
the view of \citet{tur82} that the SFR ON\,2 is physically associated
with the massive star cluster Berkeley\,87 hosting the WO type star
WR\,142.  {\new We discuss different explanations for the apparently diffuse 
X-ray emission in these SFRs. These include 
synchrotron radiation, invoked by the co-existence of strongly shocked
stellar winds and turbulent magnetic fields in the star-forming
complex, cluster wind emission, or an unresolved population of point
sources.} 
\end{abstract}


\section{Introduction} 

Recent observational advances in X-ray astrophysics have led to a
new high-energy perspective on the interstellar medium and
star-forming regions (SFRs). The X-ray point sources in SFRs comprise
pre-main sequence (PMS) stars as well as massive stars. Sometimes,
X-ray emission from deeply embedded young stellar objects (YSO) is
observed, albeit such observations remain rare \citep{pr09}.  It is
now firmly established that the shocked winds of young massive OB
stars contribute to the heating of interstellar matter up to $\lsim
10$\,MK \citep{tow03,gu08}. In older massive stellar {\nchanged
aggregates},
the combined action of stellar winds and supernova explosions results
in powerful cluster winds and associated superbubbles
\citep{cc85}. Higher temperatures (up to 100\,MK) are observed in
these objects. Hot, X-ray-emitting gas is present around massive
stars and fills the large volumes of star clusters, and, in some cases
beyond.

There is a small group of SFRs where hard,  sometimes non-thermal X-ray
emission is observed. In cases when strong fluorescent lines are seen
in the spectra, the emission is explained by the presence of a recent
supernova remnant (SNR) and its interaction with the cool dense material 
of a nearby molecular cloud \citep{tak02}. However, in some prominent 
SFRs, such as RCW\,38, the non-thermal X-ray emission cannot be easily
explained unless the presence of magnetic fields is assumed \citep{wo02}.

Magnetic fields, along with turbulence, play an important role in
star formation \citep{cr09}. Magnetic fields were directly
measured in M\,17 \citep{bt01} and Orion \citep{s98}, and are
predicted to be a common feature in SFRs \citep{fer09}. 
If accelerated particles are present in the same volume, 
synchrotron emission would naturally occur.

A region within a few arcminutes of the OH maser ON\,2 in Cygnus\,X is
an established site of ongoing massive star formation. Following the
literature we refer to this whole SFR as ON\,2 \citep{dent88, shep97}.
This region is located within the well-studied massive star cluster \ber.

One of the most interesting members of \ber\ is the WO type star
WR\,142.  Only three stars of this spectral type are known in the
Galaxy, and the WO star in \ber\ is the closest among them.  Stars 
of this spectral type represent the evolutionary stage immediately
before the supernova or $\gamma$-ray burst explosion, and drive the
fastest stellar winds among all stars. Berkeley 87 traditionally
attracts the attention of high-energy astrophysics as a potential site
of particle acceleration. Therefore, SFR ON\,2 provides an ideal
laboratory to study the interactions between {\changed{an active 
star-forming region}} and the massive star feedback.

In this paper we present our \xmm\ observations of the field
encompassing ON\,2 and \ber, and provide their analysis and
interpretation.  The paper is organized as follows.  In Section
\ref{sec:on2} we introduce the SFR ON\,2, and discuss its distance and
relation to the star cluster \ber. The membership and evolution of
\ber\ is treated in Sect.\,\ref{sec:ber87}. The analyses of \xmm\ and
\spitz\ observations are presented in Sect.\,\ref{sec:obs}. In
Sect.\,\ref{sec:pms} we briefly address the distribution of point
sources detected by \spitz.  In Sect.\,\ref{sec:ob} we consider X-rays
from massive stars in \ber.  Section \ref{sec:on2n} is devoted to the
X-ray emission from the \hii\ region GAL\,75.84+0.40. Section
\ref{sec:on2s} concentrates on the observed properties of X-ray
emission from ON\,2S, the southern part of ON\,2, while discussion on
its origin is presented in Sect.\,\ref{sec:oon2s}.  A comparison to
other SFRs is presented in Sect.\,\ref{sec:sfr}, and conclusions are
drawn in Sect.\,\ref{sec:cncl}. {\changed{In the Appendix we briefly
review the suggestions in the literature about the possible identification
of $\gamma$-ray sources in \ber.}}

\section{The massive star-forming region ON\,2} \label{sec:on2}

\begin{figure*}[!tb]
\includegraphics[width=0.99\textwidth]{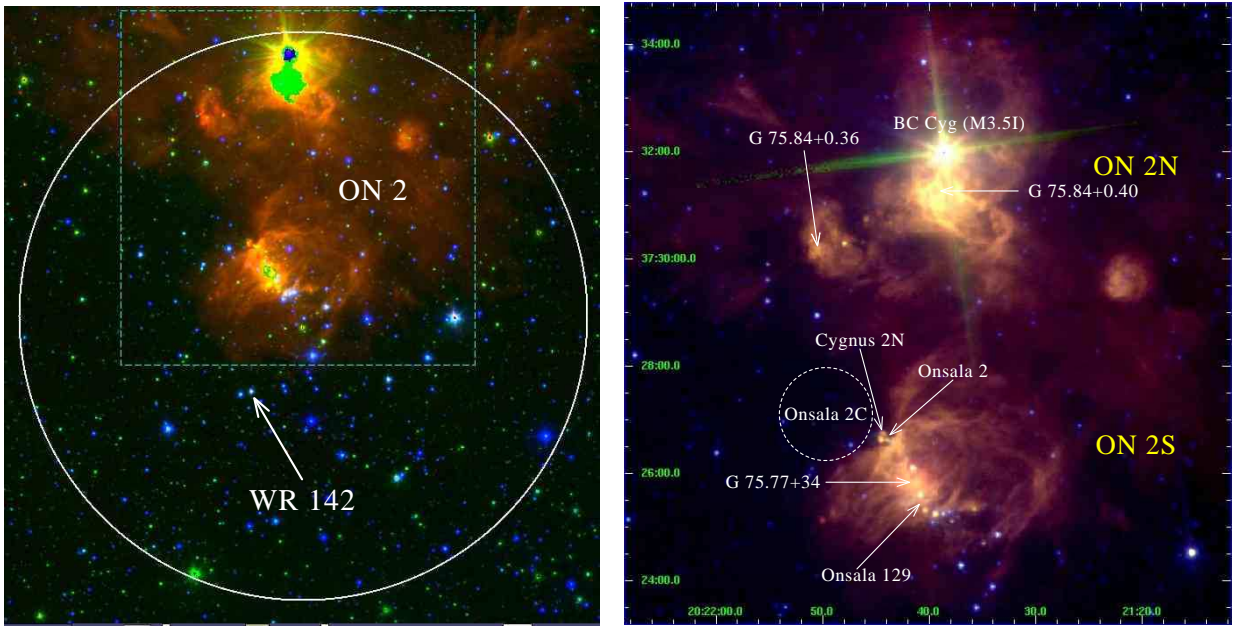}
\caption{{\changed{\it Left panel}}: Combined optical and IR image of 
Berkeley\,87 and ON\,2.  The POSS2/blue  image (Copyright Second 
Palomar Sky Survey 1993-1995 
by the California Institute of Technology) is shown in blue, the
\spitz\ IRAC channel 1 (3.6\,\mim) is in green, and channel 4 
(8\,\mim) in red.  The large white  circle represents 
the outer cluster boundary as determined  from star counts  by
\citet{tur82}. The center of the circle is at 
20$^{\rm h}$21$^{\rm m}$37$^{\rm s}$,
+37\degr\,24\arcmin\,37\arcsec\, (J2000), 
the radius is $8'$. {\changed The \ber\ cluster is fully within the \xmm\
field-of-view ($30'$). The approximate extend of the  massive star-forming 
region  ON\,2 is shown by the dashed square. The region withing the square 
is enlarged in the right panel. {\it Right panel}: Combined IR {\em Spitzer} 
IRAC (3.6\,\mim\ blue, 4.5\,\mim\ green, 8.0\,\mim\ red) image of the 
massive star-forming region ON2. The \hii\ regions described in the text 
are identified by arrows. The approximate location of the molecular 
cloud Onsala\,2C is shown as a dashed circle. Image size is 
$\approx$$12\arcmin$$\times$$11\arcmin$.} North is up, and east is left.}
\label{fig:b87}
\end{figure*}
%

\begin{figure}[!tb]
\includegraphics[width=0.99\columnwidth]{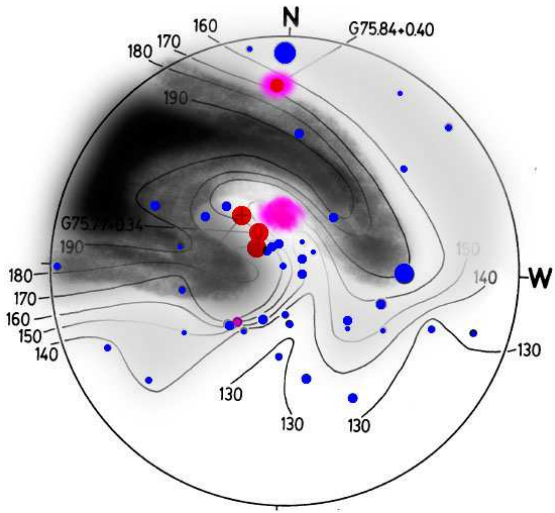}
\caption{Sketch of the possible morphology of ON\,2 based on
Fig.\,6 from \citet{tur82}. The black circle is the same as 
in Fig.\,\ref{fig:b87}. The shaded area follows contour lines
joining regions of similar space reddening (units of 0.01\,mag in 
$E_{B-V}$). The pink elliptical regions represent approximately 
the patches of hard X-ray emission. The blue dots 
represent stars in \ber, the red spots represent UC\hii\ regions. 
Not to scale.}
\label{fig:ris}
\end{figure}

Figure\,\ref{fig:b87} shows the composite DSS optical and \spitz\ IR
image of SFR ON\,2 and star cluster \ber.  A large molecular cloud
complex, Onsala\,2C, observed in CO, is intimately associated with the
whole ON\,2 complex \citep{mat86}.  In a morphological model by
\citet{tur82} based on studies of the reddening and obscuration, \ber\
sits on the western edge of the heavily obscured SFR ON\,2 and the
associated giant molecular cloud Onsala\,2C.  Figure\,\ref{fig:ris}
gives a schematic representation of the morphology of the ON\,2
region. The center of the massive star cluster Berkeley 87 is in the
middle of the figure. The molecular cloud occupies the upper left
quadrant. The sites of active star formation, as highlighted by
compact \hii\ regions, are located at the edges of a molecular cloud.

A \spitz\ IRAC image of the region is shown in
{\changed{the right panel of}} Fig.\,\ref{fig:b87}.  Following
\citet{dent88} we will distinguish between northern and southern \hii\
regions, and introduce the notations ON\,2N and ON\,2S. The northern
part, ON\,2N, comprises the \hii\ regions GAL\,75.84+0.40 and
GAL\,75.84+0.36.  ON\,2S contains the \hii\ regions Cygnus\,2N (alias
G75.78+0.34, Onsala\,2N), Onsala\,2 (alias [HLB98] Onsala\,130),
[L89b]\,75.767+00.344, and [HLB98] Onsala\,129
(Fig.\,\ref{fig:b87}). The UC\hii\ regions Cygnus\,2N and Onsala\,2
are separated by only $\approx$ 2\farcs5
\footnote {The Simbad data-base gives aliases of Onsala\,2 as
[WAM82]\,075.77+0.34, OH 75.8+0.3, [HLB98]\,Onsala 130, and
[PCC93]\,414. The coordinates of Onsala\,2 from Simbad are
\mbox{20$^{\rm h}$21$^{\rm m}$43$\fs$8,
+37\degr\,26\arcmin\,39\arcsec\, (J2000)}. However, the coordinates of
[WAM82]\,075.77+0.34 are \mbox{20$^{\rm h}$21$^{\rm m}$41$\fs$31,
+37\degr\,25\arcmin\,53.5\arcsec\, (J2000)} \citep{wam82}. This object
is $\approx$54\arcsec\ away from Onsala\,2. The positional accuracy of
\citet{wam82} is typically $5\arcsec$. Therefore, [WAM82] 075.77+0.34
{\bf is not} Onsala\,2. }.
Onsala\,2, Cygnus\,2N, and Onsala\,129 \citep{pcc93,hc96,shep97}.

\citet{shep97} studied ON\,2S in molecular lines and the 3\,mm
continuum. They detected three deeply embedded YSOs in Cygnus\,2N, one
of which is likely to be the driving engine of a molecular outflow. From
the dynamical timescales of the outflows ($\sim$30-50\,kyr) and the high
luminosities inferred for the YSOs (5000, 430, and 330\,\Lsun) they
suggest that a near-simultaneous massive star formation event occurred
in this region $\sim 10^4$\,yr ago. This support similar conclusions 
made by \citet{dent88}.

%
%

\subsection{Distance to ON\,2 and its relation to \ber } 
\label{sec:onber}

ON\,2 is located in the Cygnus region, and thus we are looking
tangentially to the Orion local spiral arm and observe numerous objects at
different distances \citep{uy01}.  \citet{rei70} estimate the distance to
ON\,2 as 5.5\,kpc, while \citet{tur82} argue that ON\,2 is located 
at the same distance as the cluster Berkeley\,87  and is
physically connected with this cluster.  \citet{mas01} obtained new
photometric measurements of stars in \ber\ and derived a distance 
of $d\approx 1600$\,pc. \citet{tur06} augmented their previous studies 
of \ber\ by a larger number of stars, and derived $d\approx 1230\pm
40$\,pc, which we adopt here.

\citet{tur82} notice that a trunk-like zone of heavy obscuration seen in
Fig.\,\ref{fig:ris} corresponds spatially with a CO clouds belonging
to the Cygnus\,X complex. In a recent study \citet{schn07} show that
the UV radiation from the clusters within the Cygnus OB1 association,
including \ber, affects the molecular cloud complex in the Cygnus\,X
south region, and note that a distance between 1.1 and 1.3\,kpc
is favored from O stars spectroscopy. This supports the likely
association between Berkeley\,87 and ON\,2.

\section{The open star cluster Berkeley\,87} \label{sec:ber87}

%
\begin{figure*}[!tb]
\centering
\includegraphics[width=0.75\textwidth]{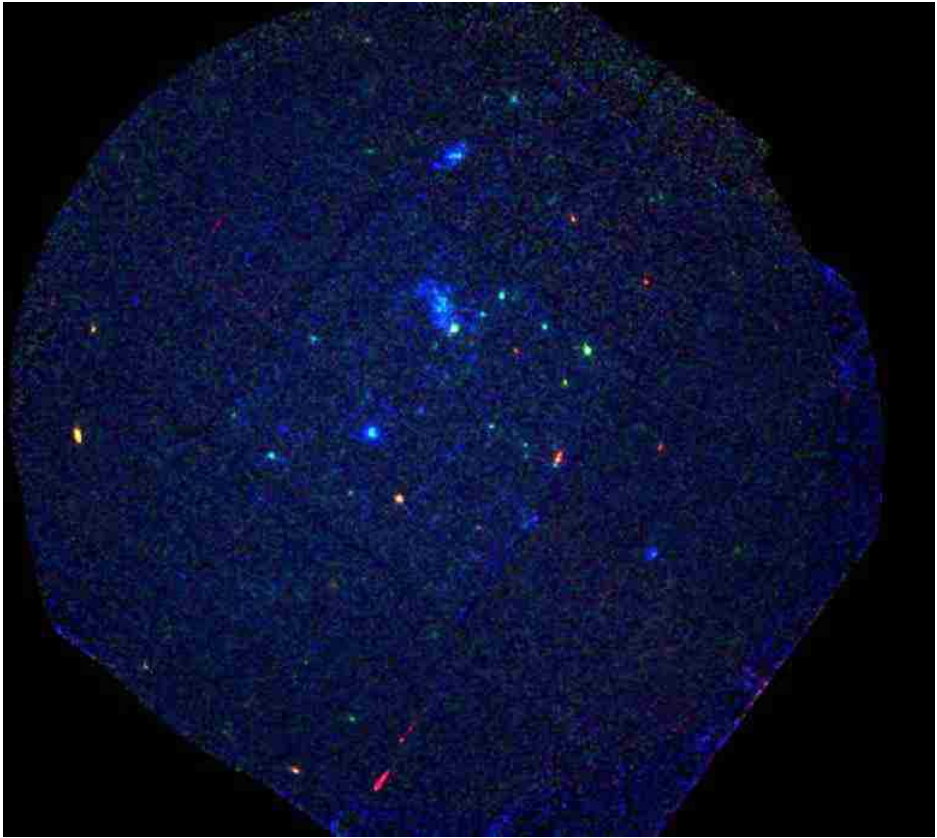}
\caption{Combined \xmm\ MOS1, MOS2, and PN image of the \ber\
in the 0.25 to 12.0\,keV band (red=0.25-1.0\,keV, green=1.0-2.5\,keV,
blue=2.5-12.0\,keV). Image size is $\approx$$30\arcmin$$\times$$30\arcmin$.
North is up, and east is left. }
\label{fig:xb87}
\end{figure*}
%

\ber\ is a relatively sparse, moderately reddened cluster, whose members
represent some of the key stages in the evolution of massive stars.
The most intriguing is a WO-type star, WR\,142. Analysis of its X-ray
observations was presented in \citet{osk09}.  The brightest member of
\ber, HD\,229059, lies slightly off the cluster core. It is a binary
system with a B1.5Iap and a lower-luminosity late-O or B0 star
companion \citep{neg04}. \citet{m87} suggests that this star is a blue
straggler.  Close to the cluster center is the peculiar variable star
V\,439\,Cyg. \citet{neg04} identify this star as having B1.5Ve
spectral type, while \citet{mas01} suggest a B[e] classification. The
bright red supergiant BC\,Cyg of M3.5Ia type \citep{tur06} is located
at the north of the cluster, close to the \hii\ region
GAL\,75.84+0.40. A possible spectroscopic binary BD+36{\degr}4032
(O8.5\,V or O8.5III) \citep{neg04,mas01} is located immediately south
of ON\,2S.

%
\begin{figure*}[!tb]
\centering
\includegraphics[width=0.9\textwidth]{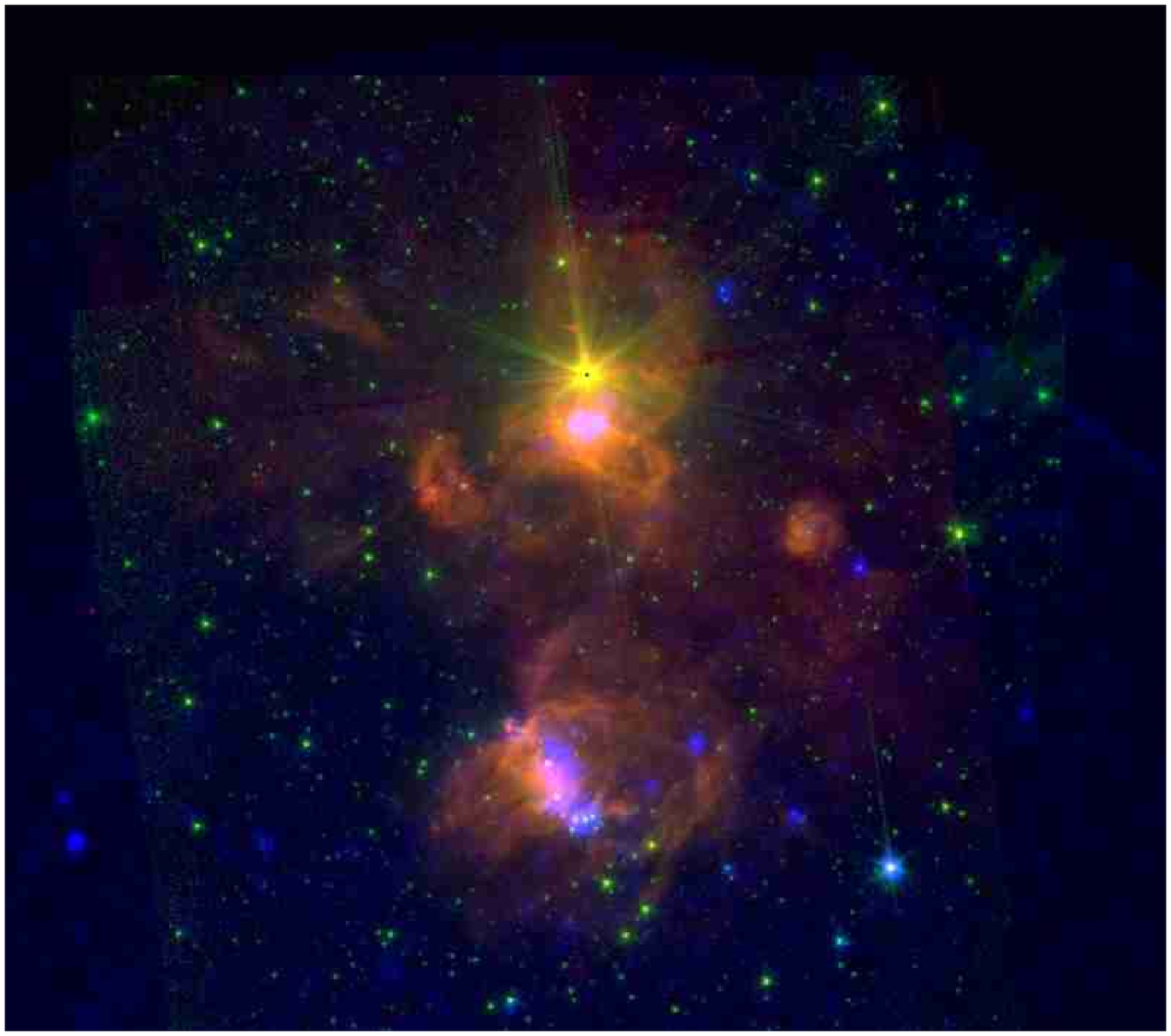}
\caption{Image of ON\,2 combined from 
\xmm\ EPIC (0.25 to 12.0\,keV band, blue) and  \spitz\ IRAC (3.6\,\mim, 
green, and  8.0\,\mim, red).}
\label{fig:irx}
\end{figure*}
%

The history of \ber\ is not well understood. The ages of OB stars, 
the WO star, and the red supergiant star are in apparent disagreement
with an evolution within a nearly coeval cluster.

\citet{vb98} found that 27\% of the young Galactic massive clusters
contain WR and {\changed{red supergiant}} (RSG) members, and that such
clusters must be older than 4\,Myr. Moreover, WC/WO stars can co-exist
with RSG stars only during a short time interval of $\sim {\rm
a~few}\times 10^5$\,yr. \citet{mas01} studied \ber\ among other
Galactic open clusters. They suggested that OB stars in \ber\ were
formed within the time span of $<$1\,Myr and are coeval with WR\,142
at 3.2\,Myr. Using evolutionary tracks of Schaller \etal\ (1992), they
estimate the initial mass of \wo\ as $M_{\rm i}=70$\,$M_\odot$.  In
the meantime, the initial mass of BC\,Cyg is $\lsim 25$\,\Msun\
\citep{lev05}. A star of such mass must be older than 6.4\,Myr when it
reaches RSG stage \citep{sch92}.

This apparent contradiction can be resolved with stellar evolutionary
models that account for rotation \citep{mm05}. We have recently found
indications that \wo\ may be a fast rotator \citep{osk09}.  In fast
rotating stars, the RSG stage will occur earlier, perhaps even during 
the H-burning stage, while the WO stage occurs later compared to the
non-rotating models. Therefore, the simultaneous presence of a RSG and a
WO star can be explained if \ber\ is $\sim$\,4-6\,Myr old.

Assuming that the most massive star in \ber\ has an initial mass
higher than 80\,\Msun\ and the universal initial mass function (IMF),
we estimate that about 30 stars with initial mass more than
10\,$M_\odot$ should be present in \ber.  This estimate is consistent
with the 22 known massive stars in \ber\ given the uncertainties
\citep{mas01}. In this case, the total mass of \ber\ {\changed{should
be}} $\approx 1200$\,\Msun. The number of low mass stars with masses
between 0.5\,\Msun\ and 3\,\Msun\ {\changed{can be estimated as}}
$\approx$\,2000.

\section{\xmm\ and \spitz\ observations of \ber.}
\label{sec:obs}

\ber\ was observed by \xmm\ during two consecutive satellite orbits
\footnote{ObsId\,0550220101, ObsId\,0550220201}. The data were merged
and analyzed using the software {\sc sas}\,8.0.0.  After the 
high-background time intervals have been rejected, the combined
exposure time of all detectors was $\approx 100$\,ks.  We followed the
standard procedure for the source detection, setting the minimum
detection likelihood to 5.  That yielded the detection of 130 point
sources as well as two regions of diffuse X-ray emission.

A combined EPIC MOS1, MOS2, and PN image of \ber\ and its surroundings
is shown in Fig.\,\ref{fig:xb87}.  Two prominent hard (blue color)
extended X-ray sources are located in the ON\,2 SFR, slightly off to 
the north from the image center.  The combined X-ray and IR images of
ON\,2 are shown in Fig.\,\ref{fig:irx}.  The extended X-ray emission
traces the eastern edge of ON\,2S region and is also observed from the
\hii\ region G75.84+0.40 in ON\,2N.

%
\begin{figure*}[!tbp]
\center
\includegraphics[width=0.99\textwidth]{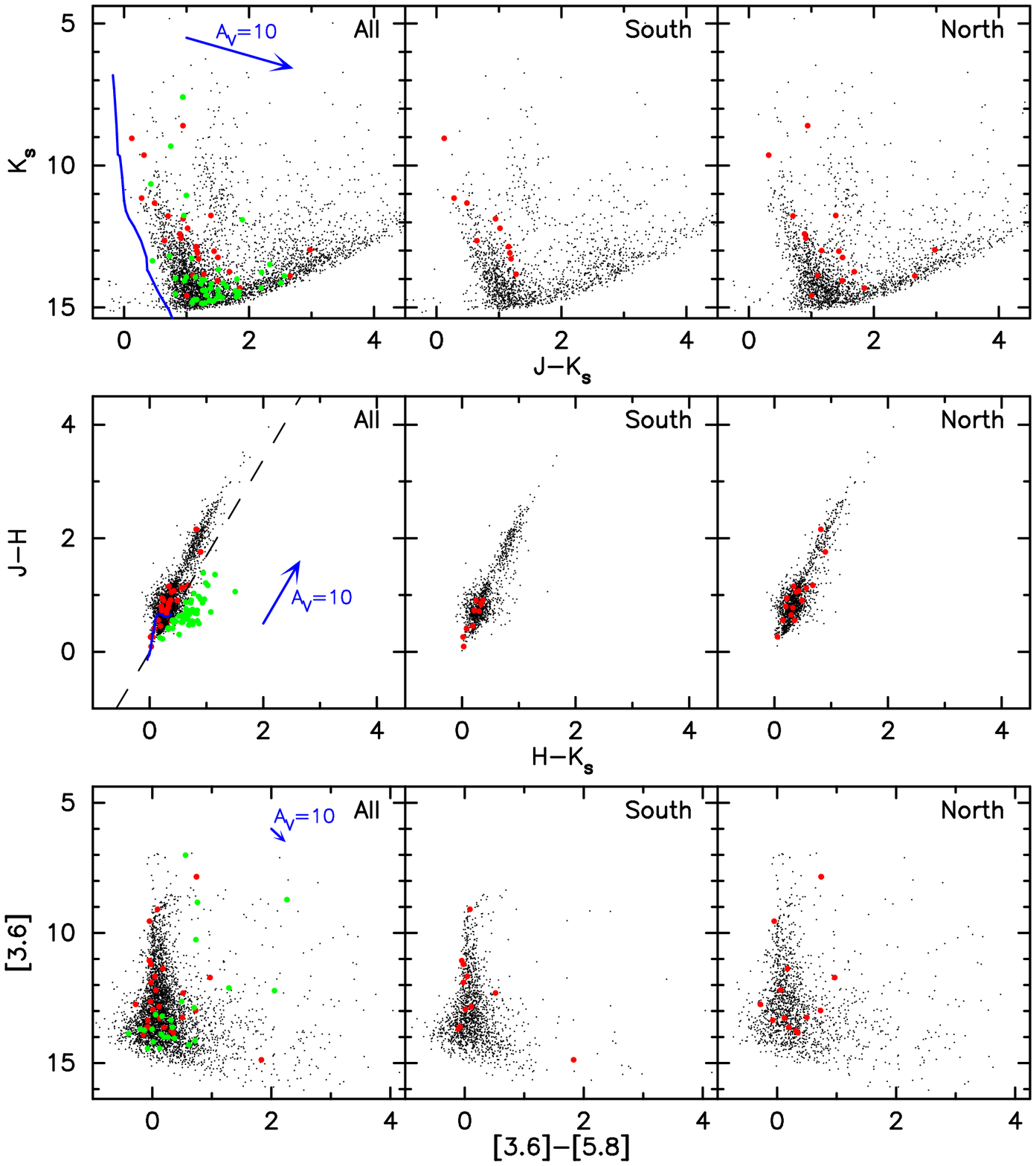}
\caption{
CMDs and CCD based on IRAC and 2MASS photometry. The panels marked
``all'' use the photometric data for the entire field, while the
panels marked ``south'' use a subset which exclude the ON\,2 SFR, and
the panels marked ``north'' correspond to the ON\,2 SFR.  In the
leftmost panels of each row the arrow indicates the reddenning vector
expected for A$_V$=10 \citep{Cetal89} and the solid lines in in the
$K{_s}$ versus $J-K_{s}$ CMD and $J-H$ versus $H-K_{s}$ CCD correspond
to the expected location of the zero-age main sequence for an assumed
distance of 1.23~kpc \citep{BB88}.  Objects plotted with a filled red
circle have a possible X-ray point source counterpart.  
{\changed{Objects plotted with green circles are more than 1-$\sigma$ 
beyond the black dashed line ($J-H=1.692 H-K$) indicating possible excess
infrared emission \citep{Aetal04}.}}} 
\label{fig:cmd}
\end{figure*}
%

The angular resolution of \xmm\ is $\lsim 6\arcsec$. To exclude the
potential confusion of extended emission with an unresolved stellar
point source, we inspected optical and {\nchanged infrared} images with
higher angular resolution.  The region around the Berkeley\,87 cluster
has been partially imaged in the mid-IR with the \spitz\ InfraRed Array
Camera \citep[IRAC;][]{Fetal04} with angular resolution of
$\lsim$2\arcsec .  The northern half of Berkeley\,87, including the
\hii\ regions GAL\,75.84+0.40 and GAL\,75.78+0.34, has observations in
all four IRAC bands (3.6, 4.5, 5.8 and 8.0~$\mu$m) but the southern half
of Berkeley\,87 is only covered in 3.6 and 5.8~$\mu$m images.  We
downloaded all applicable IRAC observations from the \spitz\ archive and
combined the basic calibrated data to form mosaic images of the entire
region using the MOPEX software package.  More information on the
instruments and pipeline processing can be found at the \spitz\ Science
Center's Observer Support
website\footnote{http://ssc.spitzer.caltech.edu/ost.}.

\section{Brief analysis of distribution of IR sources in \ber. }
\label{sec:pms}

{\changed YSOs can be identified by their IR excess, since  they are 
still surrounded by dusty disks and envelopes that absorb   stellar
light and radiate at IR wavelengths. Theoretical predictions regarding
the location  of YSOs in color-magnitude diagrams (CMDs) and color-color
diagrams (CCDs) {\nchanged can be used} to compare with observations in
order to study the YSOs.  Based on \spitz\ IRAC and MIPS observations, 
this method was recently used by \citet{gc09}  to search for YSOs in the
Large Magellanic Cloud. We use a similar method  to search for objects
with near- and mid-infrared excess in \ber.}

Aperture photometry was performed on the 3.6,
4.5, 5.8, and 8.0~$\mu$m IRAC images to obtain mid-IR flux densities
for sources throughout the \ber\ region.  The
results at each wavelength were combined for sources with $<$1\farcs 5
positional coincidence.  We also obtained near-IR ($JHK_s$) flux
densities from the Two Micron All Sky Survey Point Source Catalog
\citep[2MASS PSC;][]{Setal06} and combined these with the mid-IR
measurements, again requiring a $<$1\farcs 5 positional coincidence
for a positive match between sources detected in different bands.

\begin{figure}[!tb]
\center
\includegraphics[width=0.99\columnwidth]{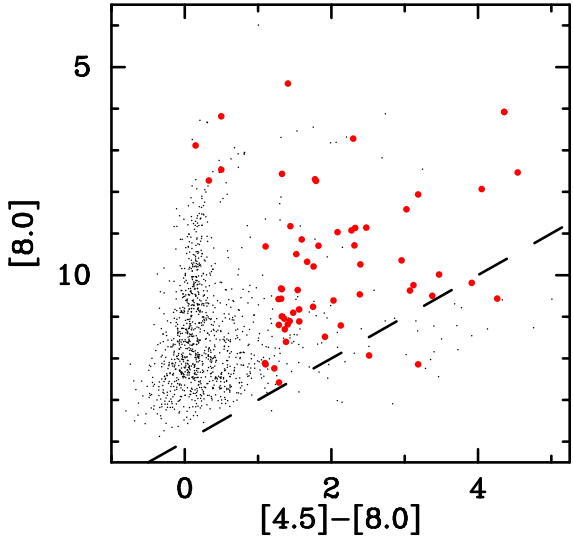}
\caption{An [8.0] versus [4.5]$-$[8.0] CMD showing all sources
in the \ber\ region.  Red points indicate sources from Figure~\ref{fig:cmd}
with [3.6]-[5.8]$>$1.0.  Background galaxies and AGN should generally 
fall below the dashed line, [8.0]$=$14$-$([4.5]$-$[8.0] \citep{Hetal06}.}
\label{fig_notGAL}
\end{figure}
%
%
\begin{figure}[!tb]
\includegraphics[width=0.99\columnwidth]{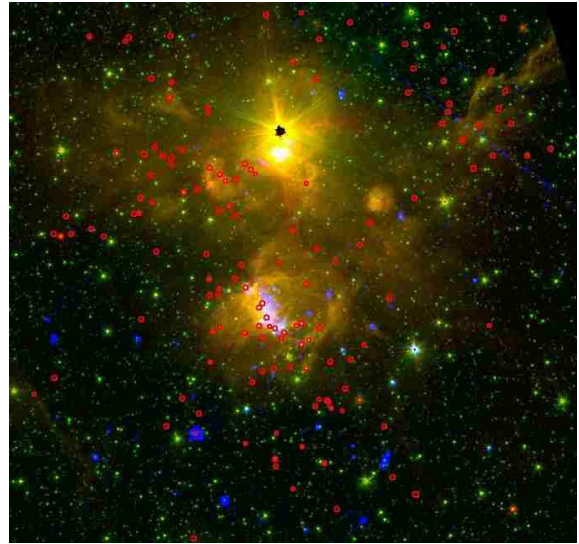}
\caption{Combined {\em Spitzer} IRAC (3.6\,\mim\ green, 4.5\,\mim\ red)
and \xmm\ EPIC (blue) image of \ber. The red regions mark the positions 
of IRAC sources with {\nchanged [3.6]$-$[5.8]$>$1.0} which may be YSOs.
Image size is $\approx$$22\arcmin$$\times$$20\arcmin$. 
North is up, and east is left.  }
\label{fig:pms}
\end{figure}
%

The resulting catalog does not have complete spatial coverage at all
wavelengths over the entire region discussed in this paper.
Specifically there are no IRAC observations at 4.5 and 8.0~$\mu$m for
declinations south of +37$^\circ$20$^{\prime}$ (the southern
and unobscured half of Berkeley\,87).  Therefore to search for
evidence for a population of YSOs and PMSs we have confined ourselves
to using the 2MASS $JHK_s$ bands and the IRAC 3.6 and 5.8~$\mu$m
bands.  In Fig.\,\ref{fig:cmd} we present color-color and
color-magnitude diagrams (CCDs and CMDs) that illustrate the
difference in the stellar population we see when we compare: (1) the
entire \ber\ region, (2) the southern, unobscured, portion of \ber\
and (3) the northern portion of \ber\ which contains the SFR ON\,2.
Specifically the sources for the entire \ber\ region are within
a box centered at 
RA=20$^{\rm h}$21$^{\rm m}$37\rlap{$^{\rm s}$}{.}7,
DEC=+37\degr\,24\arcmin\,43\farcs5, 
with $\Delta$RA=20\arcmin, $\Delta$DEC=20\arcmin, 
while the southern region is defined by a box centered at 
RA=20$^{\rm h}$21$^{\rm m}$37\rlap{$^{\rm s}$}{.}7,
DEC=+37\degr\,17\arcmin\,22\farcs5, 
with $\Delta$RA=20\arcmin, $\Delta$DEC=10\arcmin, 
(i.e. excluding the ON\,2 SFR), and the northern region is defined by 
a box centered at
RA=20$^{\rm h}$21$^{\rm m}$37\rlap{$^{\rm s}$}{.}7, 
DEC=+37\degr\,28\arcmin\,28\farcs4, with 
$\Delta$RA=20\arcmin, $\Delta$DEC=12\arcmin, 
(i.e. the ON\,2 SFR and surroundings).
While the near-IR CCDs and CMDs constructed from 2MASS data appear
similar in the north and south, the [3.6] versus [3.6]-[5.8] CMD,
constructed from the IRAC observations, reveal a higher density of
sources with red mid-IR colors (excess mid-IR emission).  After correcting 
for the area we find that for [3.6]-[5.8]$>$1.0 there are $\sim$4 times 
as many red sources in the northern region than in the south 
{\changed{(see lower panel in Fig.\,\ref{fig:cmd}).

%
\begin{figure*}[!tb]
\centering
\includegraphics[width=0.7\textwidth]{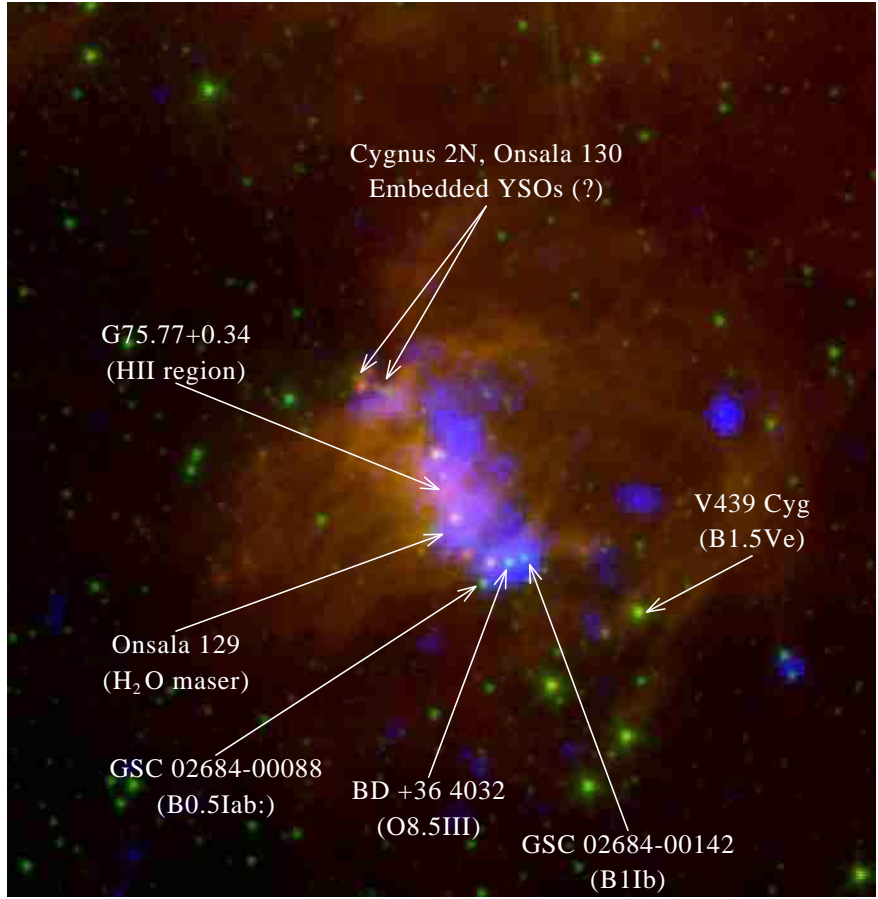}
\caption{Combined \xmm\ MOS2 (0.25 to 12.0\,keV band, blue) and 
\spitz\ IRAC (3.6\,\mim, green, and  8.0\,\mim, red) image of ON\,2S. 
The image size is $\approx 6.5\arcmin \times 6.5\arcmin$.} 
\label{fig:on2cs}
\end{figure*}
%
%

The lack of MIPS 24~$\mu$m observations over much of the region, the
incomplete IRAC coverage, and the limited sensitivity  of 2MASS prevent
a more detailed assessment as to the  true nature of these red sources. 
Nevertheless, we speculate that these are YSOs  associated with \ber. 
For instance, if we select from the $J-H$ versus $H-K_s$
CCD those 2MASS sources with near-IR execss, we find 49 sources  (green
circles in the middle panel of Fig.\ref{fig:cmd}). Only three  of these
objects with near-IR excess are among the sources with mid-IR excess
identified in the [3.6] versus [3.6]-[5.8] CMD. Thus most of the sources
with mid-IR excess escape detection with 2MASS.  

%
\begin{table*}
\caption{The OB type stars detected by \xmm\ and their parameters}
\label{tab:ob}
\centering
\begin{tabular}{ccccccc} \hline \hline
Star \rule[-2mm]{0mm}{5.25mm} &   RA  (J2000) &  DEC (J2000)  &   Sp.type   & 
$\log{\frac{L_{\rm bol}}{L_\odot}}$ & $N_{\rm H}[10^{21}\,{\rm cm}^{-2}]$  & 
$\log{\frac{L_{\rm X}}{L_{\rm bol}}}$\\ \hline
HDE\,229059 &20$^{\rm h}$21$^{\rm m}$15$\fs$37 & 
$37\degr\,24\arcmin\,31.3\arcsec $ & B1Ia	   & 5.6 & 8.6 & -7.5 \\
Berkeley 87-4  &20$^{\rm h}$21$^{\rm m}$19$\fs$25 & $37\degr\,23\arcmin\,24.3
\arcsec$ 
& B0.2III & 4.7 & 7.4 & -7.1 \\
BD~+36{\degr}4032 &20$^{\rm h}$21$^{\rm m}$38$\fs$67 & 
$37\degr\,25\arcmin\,15.5 \arcsec $& O8.5III & 5.1 & 7.8 & -7.3 \\
\hline \hline
\end{tabular}
\end{table*}

We can, however, rule out that the sources with [3.6]-[5.8]$>$1.0 are 
dominated by background galaxies. In Fig.~\ref{fig_notGAL} we present
a [8.0] vs. [4.5]$-$[8.0] CMD for all sources with available data and
find that most of them lie above the cutoff that has been  used to
separate background galaxies and AGN in other population  studies
\citep{Hetal06,KK09}.  Hence we believe that there are strong
indications that the IRAC sources in \ber\ with mid-IR excess are
dominated by YSOs.  In order to confirm and better quantify  the YSO
population throughout this region, deeper near-IR and more  complete
mid-IR observations are required.}}

To better illustrate the distribution of these candidate YSOs in ON\,2 
we plot the locations of all sources with [3.6]-[5.8]$>$1.0 in 
Fig.\,\ref{fig:pms}.  We have also searched for possible X-ray
counterparts to the 2MASS and IRAC points sources by comparing their
positions and requiring a coincidence better than 3\arcsec\ for a positive
match.  A total of 47 matches were found but only one of those matches had 
a significant mid-IR excess (see Fig.\,\ref{fig:cmd}).  The near-IR 
fluxes and colors of nearly all the sources with possible X-ray counterparts
are consistent with normal main sequence stars.

The very low X-ray detection rate of young stars in \ber\ is not
surprising. The limiting sensitivity of our \xmm\ observations is
$F_{\rm X}\approx 1\times 10^{-14}$\,erg s$^{-1}$ cm$^{-2}$. Taking
into account the average absorbing column in the direction of \ber,
$N_{\rm H}\approx 8\times 10^{21}$\,cm$^{-2}$, the unabsorbed flux for
a thermal source of X-ray emission with $kT_{\rm X}=0.8$\,keV is
$F_{\rm X}\approx 9\times 10^{-14}$\,erg s$^{-1}$ cm$^{-2}$. At the
1.23\,kpc, this corresponds to $L_{\rm X}\approx 2\times 10^{31}$\,erg
s$^{-1}$.  Studies of low-mass stars in the Orion Nebular Cluster
shown that there is a correlation between stellar age, mass, and X-ray
activity \citep{flac03}. The X-ray luminosities of a few Myr old low-
and solar-mass PMSs are lower than the detection limit of our
observations.

The CMDs shown in Fig.\,\ref{fig:cmd} provide one more indirect evidence
that the SFR ON\,2 and star cluster \ber\ are located at the same
distance. The magnitudes of sources located in the ON\,2 region do not 
differ significantly from the magnitudes of sources in southern part of 
\ber. If ON\,2 were 4 times more distant than \ber\ (see discussion in
Section\,\ref{sec:onber}), one would expect many more faint  sources in
the right panels in Fig.\,\ref{fig:cmd}. 
The spatial distribution of YSOs across the entire ON\,2N
(Fig.\,\ref{fig:pms}) confirms earlier suggestions of \citet{tur82}
and \citet{dent88} that star formation occurred nearly simultaneously
over this whole SFR.

\section{X-ray emission from massive stars in \ber. }
\label{sec:ob}

In general, all massive stars with spectral types earlier than B1.5
are X-ray emitters. \citet{tur82} identified 17 stars in \ber\ as having
early OB spectral types.  Among them only three were unambiguously
detected in our \xmm\ observation. This is because the large
interstellar extinction in the direction of \ber\ hampers the
detection of soft X-ray sources such as OB stars.

A combined X-ray and IR images of ON\,2 and ON\,2S are shown in
Figs.\,\ref{fig:irx} and \ref{fig:on2cs}. The O-type giant BD~+36{\degr}4032
(Berkeley 87-25) is located at the southern tip of the
ON\,2S. Neighboring it are two B-type stars, separated by $8\arcsec$
and 15\farcs6 from BD~+36{\degr}4032. Among these only 
the O star is detected.

All spectra in this paper were analyzed using the {\sc xspec} software
\citep{xspec}.

The X-ray spectrum of BD~+36{\degr}4032 can be fitted with a thermal
plasma model ({\em apec}) with temperature $kT_{\rm X}= 0.6\pm
0.1$\,keV, and a neutral hydrogen absorption column $N_{\rm H}=(8\pm
1) \times 10^{21}$\,cm$^{-2}$. The unabsorbed model flux is $F_{\rm
  x}\approx 2\times 10^{-13}$\,erg\,cm$^{-2}$\,s$^{-1}$, corresponding
to $\Lx\approx 4\times 10^{31}$\,erg\,s$^{-1}$. This is the X-ray
brightest massive star in \ber.

Two B-type stars, HDE\,229059 (Berkeley 87-3) and Berkeley 87-4, are
also detected in our observation. Their X-ray spectra are fitted using
a thermal plasma model with $kT_{\rm X}= 0.6\pm 0.2$\,keV corrected
for interstellar column $N_{\rm H}=(8\pm 1) \times
10^{21}$\,cm$^{-2}$.

We use the $UBV$ photometry and spectral types determined in
\citet{tur82} to derive stellar bolometric luminosities and
color excesses $E_{B-V}$. The neutral hydrogen column densities are
then estimated using $N_{\rm H}=5.0\times 10^{21}E_{B-V}$ H-atoms cm$^{-2}$
\citep{bs81} and listed in Table\,\ref{tab:ob} for the stars that
are detected by \xmm. The absorption columns inferred from the analysis
of the X-ray spectra are in good agreement with those from $UBV$
photometry.

The ratio of X-ray and bolometric luminosities for the detected OB
stars (see Table\,\ref{tab:ob}) appear to be slightly lower than the
typical value of $10^{-7}$ for OB stars \citep[e.g.][]{osk05}.  This is
because the soft X-rays are missing due to the large interstellar
absorption.  Our crude one-temperature spectral models fitted to the
observed spectra underestimate the contribution from the soft spectral
range. Bearing this in mind, the level of X-ray emission from the OB
stars in \ber\ {\changed{appears to be normal}} for stars of these spectral 
types.

\section{X-ray emission from the \hii\ region GAL\,75.84+0.40 in ON\,2N}
\label{sec:on2n}

In this section we will address the \hii\ regions in the northern part
of ON\,2 (see Fig.\,\ref{fig:b87}). While no X-ray emission is
detected from GAL\,75.84+0.36, GAL\,75.84+0.40 is a spectacular source
of diffuse X-rays. We consider it in detail below.

%
\begin{figure}[!tb]
\includegraphics[width=0.99\columnwidth]{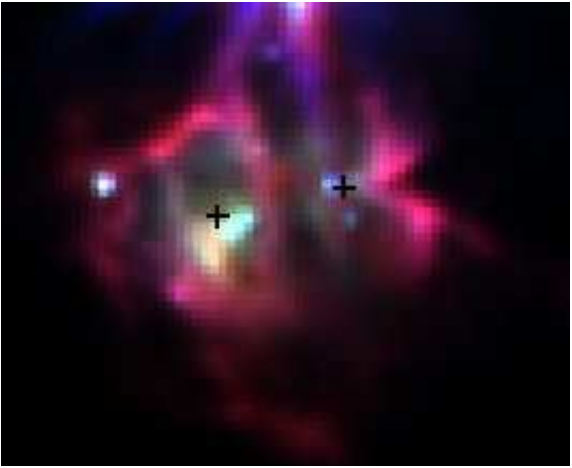}
\caption{Zoomed ON\,2N region from combined \emph{Spitzer} IRAC 3.6, 4.5, and
8.0~$\mu$m image are shown as red, green and blue, respectively.
Crosses mark the positions of source A (to the west) and source B (to the
east).  The image size is $\sim$1\farcm5$\times$1\farcm2 with north up
and east to the left.}
\label{fig:ab}
\end{figure}
%

\subsection{Morphology of \hii\ region GAL\,75.84+0.40}

{\changed
{\citet{mat73} conducted a radio survey of ON\,2. They noticed that
GAL\,75.84+0.40 has a complex morphology, and suggested that it
consists of two compact H\,{\sc ii} regions -- the first is
GAL\,75.84+0.40\,A to the east, and the second is GAL\,75.84+0.40\,B
to the west. Within each compact \hii\ region, there exists a
corresponding point source in the 2MASS PSC.  These two IR point
sources are separated by $\approx$14\farcs3 (or 0.09\,pc at
d=1.23\,kpc). Their coordinates and $K_{\rm s}$ magnitudes are given
in Table\,\ref{tab:on2n}. Corresponding sources at these positions can
be also seen in the \spitz\ images (see Fig.\,\ref{fig:ab}).}}

\begin{table}
\caption{2MASS point sources in GAL\,75.84+0.40}
\label{tab:on2n}
\centering
\begin{tabular}{cccc} \hline \hline
   \rule[-2mm]{0mm}{5.25mm} & RA (J2000) & DEC (J2000) & $m_{\rm K}$ \\
A & 20$^{\rm h}$21$^{\rm m}$37\fs 98 &
+37\degr\,31\arcmin\,15.23\arcsec & 9.628 \\
B & 20$^{\rm h}$21$^{\rm m}$39$\fs$07 & 
+37\degr\,31\arcmin\,09.27\arcsec & 9.665  \\
\hline \hline
\end{tabular}
\end{table}

In the IRAC bands source A has flux densities of 52.1$\pm$3.7~mJy and
44.1$\pm$5.0~mJy at 3.6 and 4.5~$\mu$m, respectively
{\changed{([3.6]$\simeq$9.32\,mag and [4.5]$\simeq$9.02\,mag)}},
which are also consistent with an embedded O5-6V type star. On the
other hand, for source B, we found a flux density of 194.2$\pm$7.3~mJy
at 4.5~$\mu$m ([4.5]$\simeq$7.4) which suggests a mid-IR excess if the
2MASS $K$-band source corresponds to a early type star.  Inspection of
the 2MASS and IRAC images for sources A and B reveal that both are
amid complex diffuse emission but that while source A appears to be a
single point source, source B appears elongated at all bands. We
suggest that source B may be either a multiple or its flux
measurements may suffer significant contamination from the surrounding
diffuse emission.

%
\begin{figure}[!tb]
\includegraphics[width=0.99\columnwidth]{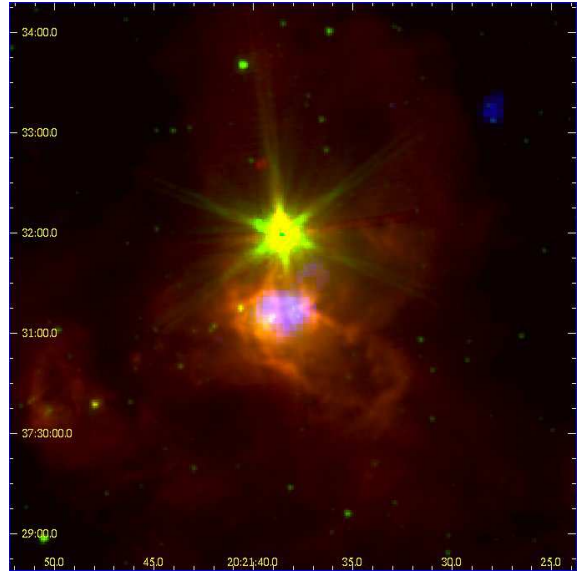}
\caption{Combined \xmm\ MOS2 (blue) and \spitz\ IRAC
3.6~$\mu$m (green) and 8.0~$\mu$m (red) images of ON\,2N.  The image
size is $\sim 5\farcm 5\times 5\farcm 5$ with north up and east to the
left.}
\label{fig:on2nx}
\end{figure}
%
%
\begin{figure}[!tb]
\includegraphics[width=0.99\columnwidth]{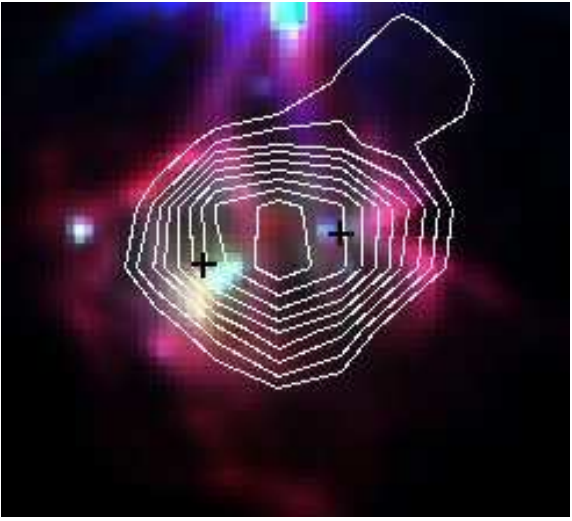}
\caption{The same as in Fig.\,\ref{fig:ab} but with contours of
X-ray emission (contours are spaced logarithmically).}
\label{fig:abx}
\end{figure}
%

\citet{gar93} presented an alternative to \citet{mat73} view on the
morphology of  GAL\,75.84+0.40. Based on VLA radio maps they was
suggested that GAL\,75.84+0.40 is an inhomogeneous shell of gas with
outer radius $\sim\,17\arcsec$, ionized by a single O-type central
star.  The highest resolution VLA 2\,cm radio continuum maps show the
two  component structure which can correspond to the bright rims of the
shell.   The "two" compact \hii\ regions  suggested by \citet{mat73} can
be the brightest parts of this shell.  {\nchanged Similar shell
morphology, with} brighten rims has been seen in planetary nebulae,
post-AGB stars, and supernova remnants.


\subsection{X-ray emission from  GAL\,75.84+0.40}

A combined X-ray and IR image of the northern part of ON\,2 is shown
in Fig.\,\ref{fig:on2nx}, while Fig.\,\ref{fig:abx} displays the IR
\spitz\ image overlaid with contours of the X-ray emission.  The
latter is extended and fills a nearly circular region centered at
\mbox{20$^{\rm h}$21$^{\rm m}$38,
+37\degr\,31\arcmin\,14\arcsec\,(J2000)}, slightly offset from the
2MASS source ``B'' (Table\,\ref{tab:on2n}). The source detection task
indicates that the X-rays are diffuse.

\begin{figure}[!tb]
\includegraphics[width=0.99\columnwidth]{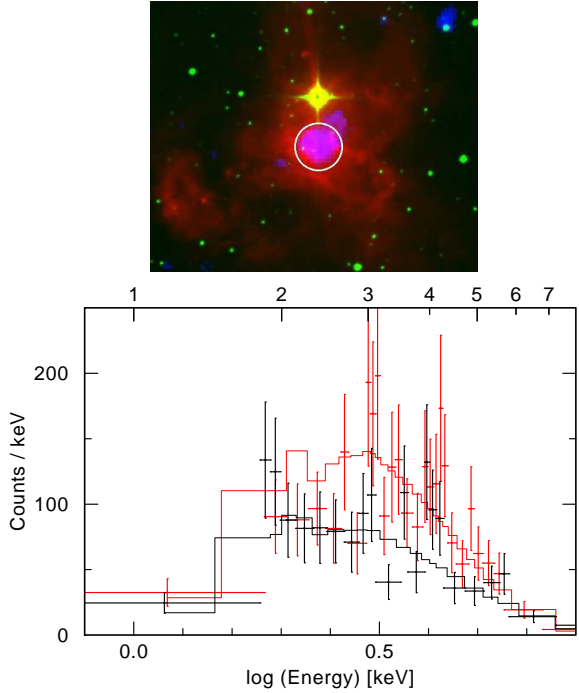}
\caption{{\it Upper panel}: Same as in Fig.\,\ref{fig:on2nx}. The white 
circle shows the spectrum extraction region. {\it Lower panel}: The PN
(red) and MOS2 (black) spectra of GAL\,75.84+0.40. The best fit
thermal plasma models plotted as solid lines for the corresponding
detectors. The parameters of the model are given in Table\,\ref{tab:onsp}. }
\label{fig:on2n}
\end{figure}

The extracted spectrum of the diffuse X-ray emission in GAL\,75.84+0.40
is shown in Fig.\,\ref{fig:on2n}. {\changed{The Cash statistics was used
to fit the spectrum by both thermal and by power-law models.}} The
spectrum has a relatively low signal-to-noise ratio and can be equally
well fitted either by thermal or by non-thermal emission. The best fit
parameters are included in Table\,\ref{tab:onsp}. The spectrum is
heavily absorbed. The absorbing column is $N_{\rm H}\approx 4.5\times
10^{22}$\,cm$^{-2}$, corresponding to $E_{B-V}\approx 9$.

%
%
\citet{tur82} determine from optical the interstellar absorption in 
the direction of GAL\,75.84+0.40 as $E_{B-V}$=1.7. Using $N_{\rm
H}=5.0\times 10^{21}E_{B-V}$\,cm$^{-2}$, the column density is 
$N_{\rm H}=8.5\times 10^{21}$\,cm$^{-2}$, i.e.\ a factor of five smaller 
than inferred from the X-ray spectral fits. Thus, we observe X-ray emission
from one or more deeply embedded sources.

\subsection{Discussion on the Origin of the X-ray Emission from 
GAL\,75.84+0.40}
\label{sec:orig}

The ionizing stars of \hii\ regions are intrinsic sources of X-ray
emission. Assuming that stars GAL\,75.84+0.40 contribute to the
observed X-ray emission, we can obtain constraints on their stellar
type. From the X-ray spectroscopy we determined $E_{B-V}\approx 9$.
Adopting $R_{\rm V}=3.1$, one obtains $A_{\rm V}=R_{\rm V}
E_{B-V}\approx 28$\,mag. The ratio between visual and $K$-band
extinction is $A_{\rm V}/A_{\rm K} = 8.9$ \citep{mon}.  From the 2MASS
Catalog, both stars in GAL\,75.84+0.40 have $m_{\rm K}\approx
9.6$\,mag, corresponding to $M_{\rm K}=-4$.  Comparing with PoWR
stellar atmosphere models \citep{wrh04}, this $K$-band absolute
magnitude corresponds to roughly an O5 type star.

\begin{table*}
\caption{Decomposing extended region filled with X-ray emission in
ON\,2S$^{\rm a}$}
\label{tab:srcx}
\centering
\begin{tabular}{cccll } \hline \hline
Region (as in Fig\,\ref{fig:cygxcon})  \rule[-2mm]{0mm}{5.25mm} & 
RA (J2000) & DEC (J2000) &
Emission & Origin \\ \hline
OB   & 20$^{\rm h}$21$^{\rm m}$38 & +37\degr\,25\arcmin\,15\arcsec &
thermal stellar wind & OB stars \\
H & 20$^{\rm h}$21$^{\rm m}$40 & +37\degr\,25\arcmin\,34\arcsec & 
? & star? UC\hii\,?\\
CN   & 20$^{\rm h}$21$^{\rm m}$43 & +37\degr\,26\arcmin\,33\arcsec & 
? & AGN? UC\hii\,?, star?\\
R    & 20$^{\rm h}$21$^{\rm m}$41 & +37\degr\,26\arcmin\,07\arcsec &
? & star? UC\hii\,?\\
D    & 20$^{\rm h}$21$^{\rm m}$41 & +37\degr\,26\arcmin\,35\arcsec &
{\new non-thermal?}   & {\new magnetic field? YSOs} \\
\hline \hline
\multicolumn{3}{l}{$^{\rm a}$ coordinates of the centers of the regions} 
\end{tabular}
\end{table*}

A number of earlier estimates for the ionizing source of GAL\,75.84+0.40
exist.  For example, \citet{dent88} used the far-infrared flux to derive 
the spectral type of the ionizing star in GAL\,75.84+0.40 to be later 
that O9.7 assuming a distance of 1~kpc. 
In contrast, \citet{mat73} found that two O8V stars were needed to 
account for the radio continuum measurements of the \hii\ region 
after adopting a distance of 5.5~kpc.  Finally, \citet{gar93}
determined that a single O6V star could explain their radio 
continuum measurements after assuming a distance of 4.1~kpc.
The only way to reconcile these earlier results with
the O5 spectral type derived from the 2MASS magnitudes and X-ray
spectroscopy is to assume that the X-ray source is embedded deeper
than the ionizing stars. Note that this conclusion is independent of
the adopted distance.

Thus, there are good arguments that the ionizing stars in
GAL\,75.84+0.40 are not the main source of the observed X-ray
emission: the X-ray source is deeper embedded, extended, and hard or
non-thermal.

The X-rays in GAL\,75.84+0.40 fill a nearly circular area, with no
obvious IR source being correlated with its center (see
Fig\,\ref{fig:abx}).  One of the possibilities to explain this can be
an embedded cluster of young stars behind the \hii\ region, which we
observe in projection, and which is the source of X-ray
emission. However, given the small linear extension of the X-ray
source (0.1\,pc) and the character of its spectrum, a cluster of
pre-main sequence stars seems to be an unlikely explanation. Albeit
some contribution from point sources is possible, we believe that a
truely diffuse component is present in GAL\,75.84+0.40.

Only a handful of star-forming regions with similar properties of
X-ray emission is presently known (see Table\,\ref{tab:hxr}).  Besides
the neighboring region ON\,2S which we discuss in the next section,
hard diffuse emission on similar spatial scales is observed in the
massive SFR RCW\,38 \citep{wo02}. An old shell-type SNR was considered
as a possible explanation. However, this explanation seems implausible
in the case of GAL\,75.84+0.40. Similar to the conclusions by
\citet{wo02} we suggest that invoking magnetic fields in the SFR {\new
may be required} to understand their hard X-ray emission. We will return to
this point when discussing the X-rays from ON\,2S in
Sect.\,\ref{sec:on2s}.

\section{The ON\,2S region} 
\label{sec:on2s}

In this and the following section we consider the southern ON\,2S
region (Fig.\,\ref{fig:b87}) and the X-ray emission detected in its
vicinity.  Fig.\,\ref{fig:on2cs} shows combined X-ray and IR images of
ON\,2S. In striking contrast to ON\,2N, the extended X-ray emission in
ON\,2S does not coincide spatially with the UC\hii\ regions and
maser sources.  Instead, the extended X-ray emission is observed to
the east of the photo-dissociation region traced by the \spitz\ images.

\begin{figure}[!tb]
\includegraphics[width=0.99\columnwidth]{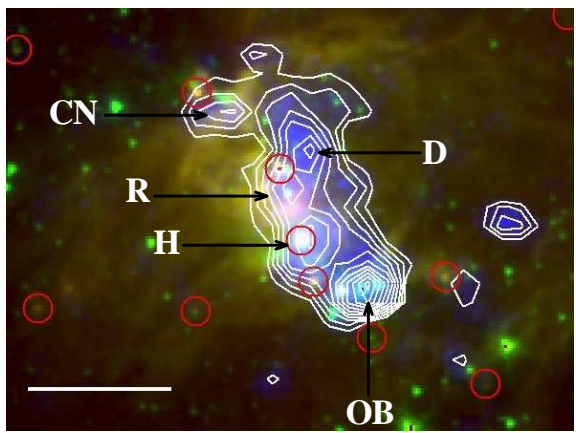}
\caption{The same as in Fig.\,\ref{fig:on2cs}, but with over-plotted
contours (log scale) of X-ray emission. The red circles are around
young stars identified from \spitz\ photometry. The regions discussed 
in the text are identified by letters and arrows. The white bar in the 
low left corner has $1\arcmin$ length.}
\label{fig:cygxcon}
\end{figure}
%

The ON\,2S region is complex and compact, and our \xmm\ data are
unable to fully resolve the individual components of the extended
region. Nevertheless, the $\lsim 4.5\arcsec$ angular resolution of the
MOS2 camera allows an investigation of the morphology of the hot gas.
Figure\,\ref{fig:cygxcon} displays the combined \spitz\ and
\xmm\ image overlaid with contours of  X-ray emission. We distinguish five
X-ray sources in ON\,2S, corresponding to the local maxima in the
contours. The coordinates of these sources are given in
Table\,\ref{tab:srcx}. The regions designated ``CN'', ``R'', ``H'', 
and ``OB'' were detected as point sources by the source detection
software. The region ``D'' is detected as a diffuse source. In the
following, we first analyze the point-sources, and then the diffuse
X-ray emission.

\subsection{X-ray emission from point sources in ON\,2S.} 
\label{sec:xpoint}

Faint X-ray emission is detected south from Cygnus\,2N (region marked
CN in Fig.\,\ref{fig:cygxcon}). Albeit this emission appears extended
in the adaptively smoothed image, the source detection software finds
a point source at this location, coinciding within 0\farcs7 with the
faint star USNO-B1.0~1274-0505781 (B1=18.08\,mag). No point source is
detected at this position in any of the \spitz\ IRAC channels.
Similarly, no continuum radio source at this position was found by
\citet{shep97}.

\begin{figure}[!tb]
\includegraphics[width=0.99\columnwidth]{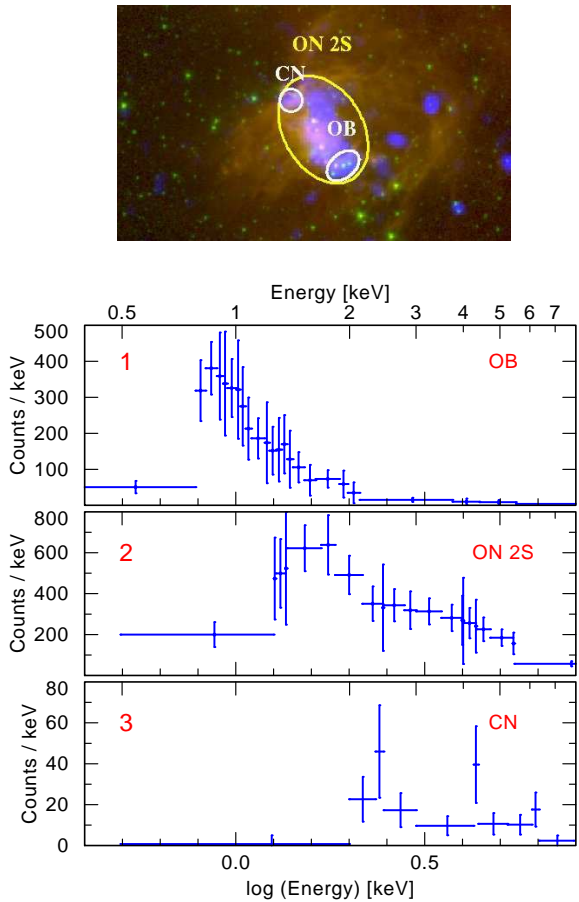}
\caption{{\it Upper panel}: Same as in Fig.\,\ref{fig:on2cs}. 
The spectrum extraction regions are shown.  {\it Lower panel}: \xmm\ PN
spectra of three regions in ON\,2S. The spectrum of the are containing 
OB-type stars is shown in plot 1.  The spectrum of whole ON\,2S is 
shown in plot 2. The spectrum of region close to Cyg\,2N 
is shown in plot 3. }
\label{fig:sp3}
\end{figure}

The X-ray emission from the point source in CN is heavily absorbed,
$N_{\rm H}\msim 2\times 10^{22}$\,cm$^{-2}$, and hard (see spectrum in
panel 3 in Fig.\,\ref{fig:sp3}). The absorbed flux is $F_{\rm
X}=(3.0\pm 1.3)\times 10^{-14}$\,erg\,cm$^{-2}$\,s$^{-1}$.  Because of
its high absorption we exclude that the source is a foreground
star. It could be either an embedded young star, or a background
object (a Galactic star or an AGN). We consider the latter possibility
as more likely, due to the lack of \spitz\ or radio counterparts.
However, on the basis of \xmm\ observations we cannot rule out the
interesting possibility that the X-ray emission south of Cygnus\,2N
originates from diffuse gas, while the close coincidence with a faint
star is accidental.

%
%
%
%
\begin{figure}[!tb]
\includegraphics[width=0.9\columnwidth]{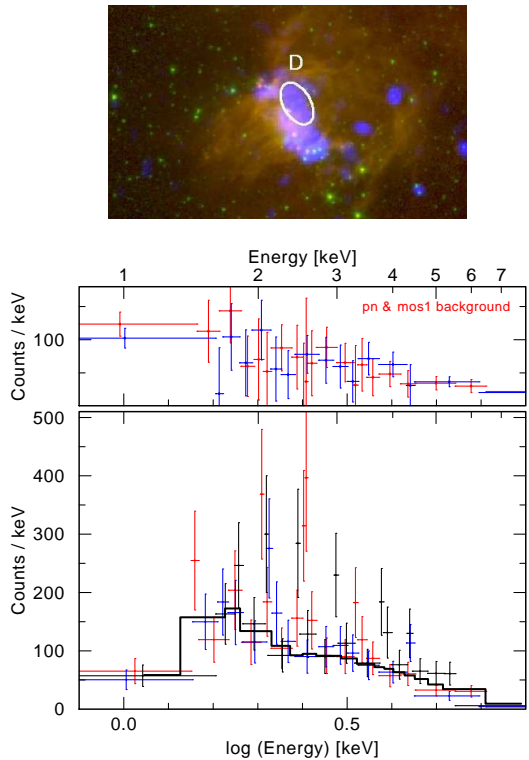}
\caption{{\it Upper panel}: Same as in Fig.\,\ref{fig:on2cs}. 
The spectrum extraction region is shown. {\it Lower panel}:
The background spectra are shown in the upper plot. 
\xmm\ PN, MOS1 and 2  spectra of diffuse emission in the region D
are shown in the lower plot. The MOS2 best fit model is shown as solid
line, parameters of the models are given in Table\,\ref{tab:onsp}. }
\label{fig:sp1}
\end{figure}

The region ``R'' is $\approx$$8\arcsec$ away from the radio source
GRS\,75.77+0.34, and $\approx$$6\arcsec$ away from a young star we
identified from \spitz\ images. It is plausible, that the radio source
and the young star are, in fact, one object. It is also plausible to
assume that the observed X-rays, at least partly, are associated with the
star.  It is interesting to note that a very red object seen only at
8$\mu$m in \spitz\ images is located in the area R.

The X-ray source in the region ``H'' (see Fig.\,\ref{fig:cygxcon})
is located some $10\arcsec$ away from the Maser\,075.76+00.34
\citep{hc96}, and 6\arcsec\ away from the optical star Berkeley\,87-83. The
X-ray emission from H is too faint to constrain its morphology, and to
extract useful spectral information.  Hence the nature of the X-ray
emission from this area remains unclear. It could be due to unresolved
young star(s) or a diffuse source.

The X-ray emission from the region ``OB'' is dominated by O star
BD~+36{\degr}4032, and was discussed in Sect.\,\ref{sec:ob}.

In Fig.\,\ref{fig:sp3} we compare the spectra extracted from the whole
ON\,2S region and from the smaller areas ``CN'' and ``OB''. It is
apparent that the spectrum of the OB region is much less affected by
absorption. The spectrum extracted from the large area is strongly
absorbed and quite hard, showing that hard X-ray emission from
embedded sources dominates the whole ON\,2S region.

\subsection{Extended hard X-ray emission from ON\,2S.}
\label{sec:ext}

{\new In the \xmm\ images} the X-ray emission from the 
region ``D'' in Fig.\,\ref{fig:sp1} appears to be truely diffuse. 
{\new We cannot fully exclude the presence of discrete sources embedded 
in the diffuse emission, albeit we did not find any such sources 
in the available multiwavelength catalogs, radio, IR and X-rays images.}
From analysis of the \xmm\ images, we
estimate that the diffuse X-ray emission occupies
$\approx$3200\,arcsec$^2$. At a distance of 1.23\,kpc this corresponds
to $\approx$0.16\,pc$^2$ (2.25\,pc$^2$ at a distance of 5.5\,kpc
respectively).

The spectrum of diffuse emission is shown in Fig.\,\ref{fig:sp1}. The
inferred absorption column density strongly exceeds the one found in
the direction of neighboring OB stars.  The quality of the spectra
is inadequate to rule out or confirm the presence of spectral
lines. There is no indication of strong Fe lines at 6.4\,keV, as
sometimes observed in SFRs \citep{tak02}.

The spectrum is quite hard and can be fitted either with a thermal or
an absorbed power-law model.  The thermal plasma model fits best with
a temperature of 200\,MK (Table\,\ref{tab:onsp}).  Such a high
temperature of interstellar gas within a small scale SFR cannot be
plausibly explained. An absorbed power-law model, $K(E[{\rm
keV}])^{-\gamma}$, where $K$ is in photons keV$^{-1}$cm$^{-2}$s$^{-1}$
at 1 keV, fits best with $\gamma=1.4\pm 0.3$. This is a plausible value
for the synchrotron emission spectrum.  It prompts us to suggest that
the relatively small volume at the west side of the SFR ON\,2S is
{\new may be filled} by synchrotron emission.

\begin{table*}
\caption{Parameters of thermal and non-thermal plasma models used to 
fit the spectra of diffuse emission in GAL\,75.84+0.40 and in ON\,2S
  (see Figs.\,\ref{fig:on2n}, \ref{fig:sp3}, \ref{fig:sp1})}
\label{tab:onsp}
\centering
\begin{tabular}{lcc}
\hline
\multicolumn{3}{c}{GAL\,75.84+0.40} \\ \hline
   & Thermal plasma   &  Non-thermal plasma \\
  &{\em tbabs*apec} & {\em tbabs*powerlaw} \\ \hline
$N_{\rm H}$\,[cm$^{-2}$]  \rule[0mm]{0mm}{3.25mm}&$(4.0\pm 0.3) \times 10^{22}$ 
& $(4.8\pm 0.6) 
\times 10^{22}$  \\
  & $kT=2.8\pm 0.3$\,keV  &  $\gamma= 2.7\pm 0.2 $ \\
$F_{\rm X}^{\rm a}$ [erg cm$^{-2}$ s$^{-1}$] & 
\multicolumn{2}{c}{$2.5\times 10^{-13}$}  \\ 
$L_{\rm X}^{\rm b}$ [erg s$^{-1}$] \rule[-2mm]{0mm}{3.25mm}& 
\multicolumn{2}{c}{$6\times 10^{31}$} \\ \hline 
\multicolumn{3}{c}{ON\,2S}  \\ \hline
   & Thermal plasma   &  Non-thermal plasma \\
  &{\em tbabs*apec} & {\em tbabs*powerlaw} \\ \hline
$N_{\rm H}$\,[cm$^{-2}$]  \rule[0mm]{0mm}{3.25mm}& $(1.3\pm 0.2)\times 10^{22}$ & 
$(1.2\pm 0.3)\times 10^{22}$ \\
  & $kT=17.7\pm 9.6$\,keV  &  $\gamma= 1.4\pm 0.3 $ \\
$F_{\rm X}^{\rm a}$ [erg cm$^{-2}$ s$^{-1}$] & 
\multicolumn{2}{c}{$2\times 10^{-13}$}  \\ 
$L_{\rm X}^{\rm b}$ [erg s$^{-1}$] \rule[-2mm]{0mm}{3.25mm}& 
\multicolumn{2}{c}{$4\times 10^{31}$} \\
\hline \hline
  \multicolumn{3}{l}{$^{\rm a}$ absorbed} \\
   \multicolumn{3}{l}{$^{\rm b}$ unabsorbed}
\end{tabular}
\end{table*}

\section{Discussion on the origin of the extended X-ray emission from
ON2\,S} \label{sec:oon2s}

The diffuse thermal X-ray emission in SFRs may result from a
collection of unresolved point sources, cluster winds, and wind-blown
bubbles.  Non-thermal emission requires the acceleration of electrons,
for which the diffuse shock acceleration (DSA) mechanism is often
invoked. In this section, we discuss the applicability of these
mechanisms to the diffuse X-ray emission from ON\,2S.

\subsection{Population of Point Sources}
\label{sec:ysos}

Unresolved X-ray emission from many low-mass YSOs present in a massive
SFR can mimic diffuse emission.  An underlying population of low-mass
stars was invoked to explain the hard extended X-ray emission observed in
the Orion and the Omega Nebula \citep{tow03,gu08}.  The known massive
YSOs in ON\,2 are 10\,000 yr old. At this age, low-mass YSOs are 
X-ray active, but may be undetected in 2MASS or IRAC images. From
Fig.\,\ref{fig:cmd} we estimate that we can see ZAMS down
to early G stars for $A_V=0$, but with just $A_V=10$ this is already
restricted to late F types. To uncover a population of low-mass
YSOs in ON\,2 high-quality, sensitive near-IR ($JHK$) imaging is needed.

Using results of \citet{flac03} obtained from the study of stars in the 
Orion Nebula, we estimate that the X-ray luminosity of a 10\,000 year old 
YSO with a mass of $0.5\lsim M/M_{\odot}\lsim 1$ is $\Lx\approx 4\times 
10^{30}$\,erg\,s$^{-1}$. Therefore, the presence of just ten such objects 
can easily explain the observed luminosity of the extended X-ray emission 
in ON\,2S.

There are, however, arguments against interpreting the emission from
ON\,2S as being solely due to an unresolved population of YSOs.
Firstly, as was discussed in Sect.\,\ref{sec:pms}, the {\em Spitzer}
data confirm earlier suggestions that star formation occurred nearly
simultaneously over this whole ON\,2 SFR.  The observed X-ray emission
is so hard, that even if it were present in the areas with higher
absorbing column we would still be able to detect it. Yet, the patch
of hard diffuse X-ray emission is spatially confined. {\changed There is no
reason to expect that low-mass YSOs are strongly clustered at   
a location that is away from the higher-mass YSOs.}

Secondly, the spectrum from an unresolved population of YSOs would be
much softer than what we observe in ON\,2S.
\citet{tow03} used {\em Chandra} observations of the Omega and the Rosetta
Nebulae to obtain a cumulative spectrum of YSOs. The absorbing columns
to YSOs in these nebulae are similar to those we derive in ON\,2S.
\citet{tow03} found that the cumulative spectrum of point sources can be 
well fitted with thermal plasma models with temperatures not exceeding
3.1\,keV. This is similar to what we infer from fitting the X-ray spectrum
of GAL\,75.84+0.40, but in ON\,2S the temperatures are much higher 
(see Table\,\ref{tab:onsp}).

{\new Therefore, while it is possible that unidentified YSOs are 
embedded in ON\,2, it appears that diffuse emission may also be present
there.}

\subsection{Shocked Stellar Wind}

The hot cluster winds filling the volumes of dense massive star
clusters are driven by stellar winds and SNe \citep{cc85}. A cluster
wind can be expected from a 4-6\,Myr old massive cluster, such as
\ber\ \citep{osk05}.

In essence, the X-ray luminosity and temperature of a wind from a
cluster with the radius $R_{\rm cl}$ depends on the input of mechanical
energy and mass. The multiwavelength observations of \ber\ do not show
any evidence of a recent SNR. {\changed{Therefore, to crudely estimate
the expected luminosity and temperature of the cluster wind from \ber, we
consider only mechanic energy and mass input produced by stellar winds. 
In \ber\ the wind energy  production is dominated by the WR star
WR\,142.  Recently, the stellar parameters of WR\,142 were updated, and
a mass-loss rate of $7\times 10^{-6}\,M_\odot$yr$^{-1}$, and wind
velocity of 5500\,km s$^{-1}$ (or 4000\,km s$^{-1}$ allowing for stellar
rotation) were derived \citep{osk09}.

To roughly estimate the energy input from other massive stars, we assume
that there are $\approx 30$ such stars in \ber\ (see
Section\,\ref{sec:ber87}). The wind velocities of O and early B stars
are in the range $\approx$1500--3000\,km s$^{-1}$, while wind velocities
of later B stars are in the range $\approx$300--1000\,km s$^{-1}$
\citep{lc99}. Mass-loss diagnostics based on comprehensive stellar wind
models that account for  macro-clumping were recently used to infer
empirical mass-loss rates of a few $\times 10^{-6}\,M_\odot$yr$^{-1}$
for typical O stars \citep{osk07, puls09}. The mass-loss rates of later
B-type stars are at least one order of magnitude smaller \citep{sea08}.
The wind-energy input from the late B and A type stars can be
neglected.  

There are eight early-type OB giants and supergiants in \ber\
\citep{mas01}. We assume that each of them has a terminal wind velocity
of 2000\,km s$^{-1}$ and a mass-loss rate of
$10^{-6}\,M_\odot$yr$^{-1}$.  Furthermore, we assume that there are 15
stars with terminal wind velocities of 600\,km s$^{-1}$ and mass-loss
rates of $10^{-8}\,M_\odot$yr$^{-1}$ (corresponding to later B type stars).
The cluster radius is 3\,pc (Fig.\,\ref{fig:b87}).

Using scaling relations of \citet{sh03}, the cluster wind luminosity is
$\Lx\approx 10^{44}\dot{M}_\ast^2 (R_{\rm cl}
\bar{V_\ast})^{-1}$\,erg\,s$^{-1}$, and the cluster wind temperature is
$T_{\rm X}\approx 15\bar{V_\ast}^2$\,MK (accounting for the 
mass-loading from "cool" matter  can lead to the somewhat lower
temperature.).  The dimension of $\dot{M}$ is \myr, $R_{\rm cl}$ is in
pc, and $\bar{V_\ast}$ is in km\,s$^{-1}$.  $\dot{M}_\ast$ is the sum of
stellar mass-loss rates in the cluster, and $\bar{V_\ast}$ is the mean
stellar wind velocity weighted by mass-loss rates. Using the wind parameters
of the stars in \ber, a very hot ($T_{\rm X}\sim 100$\,MK) but faint
($\Lx\lsim 10^{30}$\,erg s$^{-1}$) cluster wind emission can be
expected. Note, that our estimates are not sensitive to the assumed
wind parameters of the OB stars, because the kinetic energy of the WO wind
alone is an order of magnitude higher that the combined kinetic energy
input from all other stars. }}

Our \xmm\ data do not show diffuse emission filling the cluster, with
temperature peaking in the cluster center, and extending beyond its
boarders, as predicted by the analytical models. Instead, we observe
localized, small ($\lsim 0.15$\,pc$^2$) patches of diffuse emission
within the cluster.  Therefore, it is not possible to attribute the
diffuse X-ray emission from ON\,2S to the cluster wind.

If indeed the molecular cloud Onsala\,2C and the adjacent star-forming
region ON\,2 are immediate neighbors of the cluster \ber, whose
intra-cluster medium is filled with hot tenuous wind, the situation may
be analogous to the one considered by \citet{cow77} for the evaporation of
a cool cloud in a hot gas. For a spherical cloud, the solutions of
\citet{cow77} predict that the temperature rises steadily with the distance
$r$ from the cold cloud ($T\propto (1-r^{-1})^{0.4}$) and the X-ray
emission forms a "halo" around the cool cloud 
\citep[see also][]{hc09}. Broadly speaking, the thermal conduction would
lower the temperature and increase the X-ray luminosity of the hot gas
close to the interface \citep{stef08}.

There are, however, a number of arguments against attributing the
X-ray emission from ON\,2S to the thermal evaporation of cool cloud
material.  The observed X-ray temperature at the interface between
cool and hot gas appears to be unreasonably high ($\sim
200$\,MK). Moreover, the diffuse X-ray emission is localized in small
areas, while the interface between cool and hot gas in
\ber\ extends over parsecs (see sketch in Fig.\,\ref{fig:ris}).

Furthermore, as noticed e.g. by \citet{wang06}, the presence of the
Fe-line complex at $\approx 6.4-6.7$\,keV is a good indicator that
cool cloud material is involved in the generation of
X-rays. \citet{wang06} propose a cluster-cloud collision scenario,
where the emission in Fe lines traces the shocked cloud gas.
{\changed{This line complex is very prominent in the Arches and the
Quintuplet cluster spectra. Using the spectral parameters of the Fe-line
complex obtained in \citet{wang06} we conclude that if Fe emission
of comparable strength were present in ON\,2S it would be
noticeable even in our low S/N spectra. However,}} there is no indication 
for the presence of the Fe line complex in the X-ray spectra of
ON\,2S.

Thus, taking into account {\it i)} the spatial distribution of diffuse
emission, {\it ii)} the very high temperature needed to fit a thermal
model to the observed spectrum, and {\it iii)} the absence of
fluorescent iron lines in the spectra, we conclude that the diffuse
X-ray emission is {\em not} a consequence of the interaction of
the cluster wind from \ber\ with the cool cloud Onsala\,2C.

Three massive stars (two BI and one OIII) are located just south from
the area filled with extended X-ray emission (Fig.\,\ref{fig:on2cs}).
The local interaction between the winds of these three stars may, in
principle, lead to the heating of nearby regions.  This scenario,
however, is not confirmed by the data. As discussed in
Section\,\ref{sec:xpoint}, the X-ray luminosity and spectrum extracted
from the region around these three OB stars is dominated by X-ray
emission of BD+36{\degr}4032 and does not require any additional
sources such as interactions between stellar winds.

\citet{pol91} reported the detection of strong diffuse emission around 
WR\,142 in the optical. They argue that this diffuse emission likely
originates in a supersonic flow centered on WR\,142, and can be
considered as evidence of a hot bubble around this star.  Although
theoretically expected, there is a dearth of detected diffuse X-ray
emission from wind-blown bubbles around WR stars
\citep{chu03,wr05}.  The only two detected hot bubbles show a
limb-brightened morphology and are extended on the scale of parsecs. The
size of a hot bubble around WR\,142, estimated using the classical work
by \citet{we77}, should be  $\approx 40$\,pc or $\lsim 1.5\degr$ at the
distance of \ber. It should be noted, however, that \citet{we77}
considered the case of an isolated star in a uniform medium, i.e.\
conditions that are clearly not valid for WR\,142.  A superbubble and
super-shells can be expected around rich OB associations
\citep{chu03}, such as Cyg~OB\,1, which comprises star clusters
Berkeley\,86, 87, IC\,4996, and NGC\,6913. However, these structures
are extended over large scales of $\sim 10^2$\,pc. The location of
\ber\ in the complex region of the Cygnus\,X superbubble makes the
detection of an individual bubble around \ber\ even more difficult.

Thus we conclude that stellar winds cannot be the main reason for the
observed {\new apparently diffuse emission}.

\subsection{Synchrotron Emission in Random Magnetic Fields}

{\new Diffuse hard X-ray emission  with a power-law spectrum can be
produced by the synchrotron mechanism.} The latter requires the presence
of energetic particles and magnetic fields.

It has been argued that the shocks in the cluster winds and/or
colliding wind binaries efficiently accelerate electrons and protons
to relativistic energies \citep{byk01,ql05,pd06}.  The model
calculations by \citet{bed07} show that particles up to TeV energies
should be present in \ber. These particles should be advected from the
cluster on the time scale of $\sim 10^3$\,yr.  A simple scaling
relations to estimate the expected surface brightness of the
synchrotron emission from the cluster of massive stars were derived by
\citet{ql05} and applied to the cluster in the central parsec of the
Galaxy.  \ber\ is less rich than the Galactic Center cluster, but it
contains WR\,142, a WO star that alone has the wind kinetic energy
$\approx 1.5\times 10^{38}$\,erg s$^{-1}$, making it comparable to the
whole of the Galactic Center cluster.  The small number of OB stars in
\ber\ results in the order of two lower UV luminosity, thus the
cooling of particles by inverse Compton process is less significant.
According to the scaling relations of
\citet{ql05}, the surface brightness of \ber\ in $\gamma$-rays is
$\sim 100$ lower than that of the Central Cluster.

\begin{table*}
\caption{High-mass SFRs where hard diffuse  X-ray emission was
detected$^{\rm a}$}
\label{tab:hxr}
\centering
{\footnotesize
\begin{tabular}{lcccccccc}
\hline
\hline
Region \rule[-2mm]{0mm}{5.25mm}
    & \Lx\ & $N_{\rm H}$        & Size & Distance & \multicolumn{2}{c}{Model} 
& Comment$^{\rm c}$ & Reference \\
    & [$10^{32}$\,erg s$^{-1}$] & [$10^{22}$cm$^{-2}$] & [pc] & [kpc] 
& thermal$^{\rm b}$ & power law & & \\
    &                           &                      &      & 
& [keV]             & $\Gamma$  & & \\ \hline
Orion       & 0.6   & 0.1    &  2  & 2.6 & 0.4 & 
&  point sources (?) & (1) \\ 
ON\,2\,S        & 0.4   & 1.5    & 0.1 & 1.2 & 18 & $\approx 1.4$ 
& {\new synchrotron (?), point sources} & (2) \\
GAL\,75.84+0.40 & 0.6   & 4.6    & 0.1 & 1.2 & 3  & $\approx 2.6$ 
&  point sources (?)  & (2) \\
RCW\,38     & 1     & 1      & 1.5 & $\approx$2 & 1.7 & $\approx 2.8$ 
&        & (3) \\
NGC6334     & 0.1-5 & 0.5-10 & 1   & 1.7        & $>1$& 
&        & (4)  \\
Westerlund\,1 & 300 &  2     & 10  & 5          & $\msim 3$& 
& IC (?) & (5) \\
NGC 3603   & 200    & 0.7    & 4   & 7          & 3        & 
&        & (6) \\
Arches     & 200    & 10     & 3   & 8.5        & 5.7      & 
&        & (7)  \\ 
Sgr B2     & 9      &  40    & 0.2 & 8.5        & 10       & 
&        & (8) \\ 
W49A       & 30     &  50    & 0.3 & 11.4       & 7        & 
&        &  (9) \\ 
LMC\,30\,Dor\,C &   $10^4$   & 0.1 & 10  & 50   &          & 
$\approx 2.5$ & superbubble &
(10) \\
LMC\,N11  & $1.5\times 10^3$ & 0.5 & 10  & 50 & & $\approx 1.7$ & 
superbubble & (11) \\ 
LMC\,N51D & $4\times 10^3$   & 0.03&     & 50 & & $\approx 1.3$ & 
superbubble & (12) \\
\hline
\multicolumn{9}{l}{$^{\rm a}$All numbers in 
this Table are approximate. The  reader is urged to consult
the original publications describing these complex}\\
\multicolumn{9}{l}{objects and
their observations in detail.}\\
\multicolumn{9}{l}{$^{\rm b}$ Temperature of the hardest component 
in multi-temperature spectral fits}  \\
\multicolumn{9}{l}{$^{\rm c}$ Comment on the properties of suspected 
non-thermal X-ray emission, if any}  \\
\multicolumn{9}{l}{(1) \citet{gu08}; (2) this work; (3) \citet{wo02}:
(4) \citet{ez06}; (5) \citet{mu03}; } \\
\multicolumn{9}{l}{ (6) \citet{mof02}; (7) \citet{yz02}; (8) \citet{tak02}; 
(9) \citet{ts06}; }\\
\multicolumn{9}{l}{ (10) \citet{bam04}; (11) \citet{mad09}; 
(12) \citet{coo04}; }\\
\end{tabular}
}
\end{table*}

It is interesting to note that the winds from the dense group of three
OB stars south of the ON\,2S (see sketch in Fig.\,\ref{fig:ris}) may
provide an additional source of particles accelerated {\em in situ}
nearby to ON\,2S.

To estimate at which distance from the stars the winds terminate and
shock the \hii\ gas, we consider a typical O star wind density profile
which declines with distance as $r^{-2}$. The wind density close to
the surface of an O star is $\sim 10^{10}$\,cm$^{-3}$.  For the
typical ambient matter density, $\sim 1$\,cm$^{-3}$, the stellar wind
pressure would be equal to the ambient matter pressure roughly at the
distance of the patch of diffuse X-ray emission (i.e 0.3\,pc away from
OIII star BD+36{\degr}4032).


Recently, \citet{byk08} addressed the effect of a random magnetic
field on synchrotron emission.  In was shown (in the context of young
supernova remnants) that prominent localized structures can appear in
synchrotron maps of extended sources with random magnetic fields, even
if the particle distribution is smooth. The bright structures  originate
as high-energy electrons radiate efficiently in local
enhancements of the magnetic field. The size of the "patch" of hard
X-ray emission in ON\,2S is about 0.1\,pc$^2$. This could be relevant
in the framework of the Bykov \etal\ model if the shock velocity is a
few$\times 10^3$\,km\,s$^{-1}$ (Bykov A., priv. communication).  This
is a plausible number for a stellar as well as for a cluster wind.

Magnetic fields, along with turbulence, play a leading role in star
formation \citep{cr09}.  The field strengths (50-700\,$\mu$G) of an
order, or even two, higher than the field in the diffuse ISM are
detected in SFRs, moreover the field geometry is shown to be far from
uniform \citep{s98}.  These measurements provide strong support to 
the idea that magnetic fields of similar strengths can be present 
in other, if not all, SFRs, with ON\,2S being no exception.
\citet{fer09} reviewed several recent observational studies of the
relationships between magnetic fields, stellar feedback, and the
geometry of \hii\ regions.  The observations reveal that magnetic
field lines can be preferentially aligned perpendicular to the long
axis of the quiescent cloud before stars form. After star formation
and push-back occurs, ionized gas will be constrained to flow along
the field lines and escapes from the system in directions
perpendicular to the long axis. Wave motions may be associated with
the field and so could contribute a turbulent component to the
observed line profiles.

It appears that all ingredients required to produce a small bright
patch of synchrotron radiation in the X-ray image  may be present in
ON\,2S. The particles accelerated up to TeV energies may result from
the shocked cluster winds and/or from the YSO outflows. Sufficiently
strong turbulent magnetic  fields may be present in the vicinity of star
forming regions. In addition, the presence of clumps of matter would
enhance the surface brightness of the radiation \citep{byk08}, and
such clumps of matter are present in the radio maps of the region 
\citep{shep97}. {\new To summarize, a synchrotron emission may be 
expected from the region where the strongly shocked stellar winds and 
turbulent magnetic fields co-exist.}

\section{Comparison with other SFRs}
\label{sec:sfr}
%
In Table\,\ref{tab:hxr} we expand the list compiled by  \citet{ts06} to
summarize the properties of high-mass star forming regions with detected
hard diffuse X-ray emission.

It can be immediately seen that the properties of SFRs in
Table\,\ref{tab:hxr} are quite diverse. It appears that three broad
categories can be distinguished.  The superbubbles around large clusters
and associations of massive stars belong to one of these categories. 
Large spatial scale (10--100\,pc), high luminosity, and often limb
brighten morphology are characteristic for objects such as 30\,Dor\,C,
LMC\,N11, LMC\,N51D.  The star clusters blowing cluster winds belong to
the second category. The examples include Westerlund\,1, the Arches, and
NGC\,3603. Although their X-ray emission is hard, it can, usually be
fitted with hot ($T_{\rm X} \lsim 30$\,MK) thermal plasma. The diffuse
X-rays fill the cluster interior on the scale of a few pc. In the third
group, where, as we believe ON\,2 belongs, small scale areas ($\sim
0.1$\,pc) are filled with diffuse X-rays. In such SFRs as ON\,2,
RCW\.38, WR\,49A, or Sgr\,2, the hot ($T_{\rm X} \msim 100$\,MK) or
non-thermal plasma is found in immediate vicinity of UC\hii\ regions.
While \citet{tak02} and \citet{gu08} suggest that hard X-rays may
originate from unresolved populations of low-mass pre-main sequence
stars, we suggest, that at least in some objects, the non-thermal
emission {\new may be expected from} interactions between magnetic
fields and the particle accelerated in shocked stellar winds.

It is interesting to note that the distribution of hard extended X-ray
emission in the two closest SFRs -- Orion nebula and ON\,2S -- is
somewhat similar: diffuse X-ray emission filling a cavity seen in the
IR images, slightly offset from a small group of massive stars.  The
X-ray spectrum, however is noticeably harder in the case of ON\,2S.

\section{Concluding remarks}
\label{sec:cncl}

Throughout the paper we adopted the distance $d=1.23$\,kpc to both
the young massive star cluster Berkeley\,87 and the SFR ON\,2.
This assumption has allowed us to propose physical interaction between
the radiative and mechanical feedback from OB stars in Berkeley\,87
and the observed diffuse X-ray emission in the regions of active star
formation.

In the majority of papers published on the SFR ON\,2, a distance in
excess of 4\,kpc is adopted. In this case, there would be no physical
connection between the massive star cluster and the SFR.  The X-ray
luminosities and the areas filled with X-ray emission would be
factor of $\msim$16 larger. The results of our spectral analysis
would, however, not been affected.

To summarize, we have conducted \xmm\ observations of the massive 
star-forming region ON\,2. The observations and their subsequent
analysis have shown:

\medskip
\noindent
1. Diffuse X-ray emission on a cluster scale, which might have been 
expected to result from the cluster wind and wind blown bubble in \ber,
was not detected.

\medskip 
\noindent
2. From a literature search we do not confirm previous reports of 
$\gamma$-ray emission from \ber.

\medskip
\noindent
3. The northern (ON\,2N) and the southern (ON\,2S) parts of the star
forming complex ON\,2 are bright sources of diffuse X-ray emission.

\medskip
\noindent
4. X-ray images of ON\,2N show that X-ray emission fills the interior
of the compact \hii\ region GAL\,75.84+0.40. This emission is diffuse and
strongly absorbed, indicating that it originates from deeply embedded
sources.  We rule out the ionizing stars of the \hii\ region as its origin, 
and speculate that it can result from an embedded cluster of young low mass 
stars.

\medskip
\noindent
5. In ON\,2S the extended X-ray emission traces the
eastern edge of this SFR. This extended emission consists of point
sources superimposed on or immersed in diffuse emission.

\medskip \noindent 
{\new 6. We discuss different possible scenarios to
explain the apparently diffuse emisison from ON\,2S, such as cluster
wind, synchrotron radiation, and unresolved point sources. We favor the
last two options as the most probable explanations.}  

\medskip \noindent
{\em Note added in proof}.
New {\em Chandra} X-ray telescope observations of a part of \ber\
became public after this article had been accepted.  The {\em Chandra}
exposure is less sensitive than the \xmm\ observations discussed in the
present paper, but has a superior angular resolution of $\lsim 1''$. The
new {\em Chandra} images reveal three discrete sources in the region
termed D in Figs.\,\ref{fig:cygxcon}, \ref{fig:sp1}. According to a first
analysis, the {\em Chandra} spectra of these three point sources
correspond to strongly absorbed thermal emission with $N_{\rm H}\approx
3\times 10^{22}$\,cm$^{-2}$ and $kT\approx 2$\,keV. Such parameters are
usual for YSOs as discussed in Section\,\ref{sec:ysos}. Hence the new data
indicate that unresolved point sources contribute to the apparently
diffuse emission from ON\,2S detected by \xmm, confirming what we have
discussed in Section\,\ref{sec:oon2s} as one possibility. However, the sum
of the fluxes from the discrete sources is about three times smaller
than the flux from the whole region D as measured with \xmm, still
leaving open the question about the true nature of the major part 
of extended emission observed with \xmm. 

\section*{Acknowledgments}   
Based on observations obtained with \xmm, an ESA science mission with
instruments and contributions directly funded by ESA Member States and
NASA. {\nchanged This research used observations obtained with the
\spitz\ Space Telescope, which is operated by the Jet Propulsion
Laboratory, California Institute of Technology under a contract with
NASA.} The Second Palomar Observatory Sky Survey (POSS-II) made by the
California Institute of Technology with funds from the National Science
Foundation, the National Geographic Society, the Sloan Foundation, the
Samuel Oschin Foundation, and the Eastman Kodak Corporation was used in
this work. This research has made use of NASA's Astrophysics Data System
Service and the SIMBAD database, operated at CDS, Strasbourg, France.
The authors are grateful to A. Bykov and M. Pohl for the insightful
discussions.  {\changed{The useful and constructive comments of the
referee greatly helped to improve the manuscript.}}   Funding for this
research has been provided by NASA grant NNX08AW84G (Y-HC and RI) and
DLR grant 50\,OR\,0804 (LMO).

\appendix
\section{On the absence of detected $\gamma$-ray emission from \ber\ }
\label{sec:gamma}

It has long been realized that young massive stellar associations
can be potential sources of cosmic- and $\gamma$-rays, where the
acceleration of particles occurs in the shocks produced by
turbulent interactions of supersonic stellar winds
\citep{mon79,cas80,ces83}.  \ber\ is a testbed for the models of
$\gamma$-ray production due to its proximity, relatively high space
density of massive stars, and, chiefly, due to the cluster membership
of WR\,142 -- the star with perhaps the most powerful stellar wind in
the Galaxy.

However, our careful check revealed that in contrast to the
reports in the literature, the high-energy emission from \ber\ was not
previously observed.

We were not able to confirm the detection of diffuse X-ray emission from
\ber\ by {\em EXOSAT}  as referenced by  \citet{bed07}. \ber\ was observed
$\sim 1\deg$ off-axis. There is no reference in the literature to the
detection of \ber\ or any object within its boarders by {\em EXOSAT}.
Similarly, no detection of \ber\ by {\em ASCA} was ever reported.
\citet{rob02} report the identification of an ASCA X-ray source
coincident with the EGRET $\gamma$-ray source 2CG\,075+00,
GeV\,J2020+3658 or 3EG\,J2021+3716 as the young radio pulsar
PSR~J2021+3651, which is located at a distance of $\approx 10$\,kpc. The
pulsar and its associated pulsar wind nebula G75.2+0.1 are $\approx
28'$ away from the outer border of \ber.

\citet{bed07} considered two {\em EGRET} sources, 3EG\,J2016+3657
and ``3EG\,J2021+4716'' , as related to \ber\ cluster. The elliptical
fits to the 95\% confidence position contours of all 3EG EGRET sources
are presented in \citet{m01}.  The smallest separation of the outer
border of \ber\ from 95\% confidence position contours of
3EG\,J2016+3657 is $\approx 30'$. No source ``3EG\,J2021+4716'' is
present in the 3EG catalog \citep{har99}.

\citet{a06} report that the observations of  \ber\ region yielded only
upper limits that are a factor of $\lsim 3$ lower than TeV fluxes expected
from an OB stellar association. The Milagro group \citep{ab07} reported
the detection of an extended  TeV source MGRO J2019+37, but their
analysis favors its associations with the PSR~J2021+3651 and its 
pulsar wind nebula  G75.2+0.1.

From our literature search, we conclude that prior to our \xmm\
observations, there were no firm detection of X-ray diffuse sources 
from \ber. We also conclude that there are no $\gamma$-ray source 
identifications within \ber.

\appendix

\end{document}